\documentclass[journal]{IEEEtran}
\usepackage{amsmath,graphicx}
\usepackage{blindtext}
\usepackage{setspace,tabularx}
\usepackage[tight,footnotesize]{subfigure}
\usepackage{times}
\usepackage{adjustbox}
\usepackage{algorithm}
\usepackage[noend]{algorithmic}
\usepackage{amsfonts}
\usepackage{amsmath}
\usepackage{amssymb}
\usepackage{array}
\usepackage{boxedminipage}
\usepackage{cite}
\usepackage{color}
\usepackage{cuted}
\usepackage{floatflt}
\usepackage{float}
\usepackage{graphics}
\usepackage{graphicx}
\usepackage{indentfirst}
\usepackage{latexsym}
\usepackage{hyperref}
\hypersetup{colorlinks=true,linkcolor=black,filecolor=black,urlcolor=black,citecolor=black}
\usepackage{multirow}
\usepackage{picinpar}
\usepackage{placeins}
\usepackage{psfrag}
\usepackage{soul}
\usepackage{subfigure}
\usepackage[usestackEOL]{stackengine}
\usepackage{theorem}
\usepackage[flushleft]{threeparttable}
\usepackage{wrapfig}
\usepackage{mathtools}
\usepackage{romannum}
\usepackage{graphicx}
\usepackage{enumitem}
\usepackage{colortbl}
\usepackage{amsmath,graphicx}
\usepackage{amsmath,amssymb,amsfonts}
\usepackage{graphicx}
\usepackage{textcomp}
\usepackage{xcolor}
\usepackage{mathrsfs}

\graphicspath{{./fig/}}
\theoremstyle{break}

\newcommand{\ignore}[1]{ }

\newcommand{\figlbl}[1]{\label{fig.{#1}}}
\newcommand{\figref}[1]{Fig.~\ref{fig.{#1}}}

\newcommand{\seclbl}[1]{\label{sec:{#1}}}
\newcommand{\secref}[1]{Section~\ref{sec:{#1}}}
\newcommand{\eqnlbl}[1]{\label{eqn:{#1}}}
\newcommand{\eqnref}[1]{Equation~(\ref{eqn:{#1}})}

\newcommand{\beq}{\begin{equation}}
\newcommand{\eeq}{\end{equation}}

\setstcolor {red}

\title{
PaReNTT: Low-Latency Parallel Residue Number System and NTT-Based Long Polynomial Modular Multiplication for Homomorphic Encryption
}
\author{Weihang Tan$^*$,~\IEEEmembership{Student Member,~IEEE}; Sin-Wei Chiu$^*$,~\IEEEmembership{Student Member,~IEEE}; Antian Wang,~\IEEEmembership{Student Member,~IEEE};
 Yingjie Lao,~\IEEEmembership{Senior Member,~IEEE}; and Keshab K. Parhi,~\IEEEmembership{Fellow,~IEEE}% <-this % stops a space 
\thanks{Weihang Tan, Sin-Wei Chiu, and Keshab K. Parhi are with Department of  Electrical and Computer Engineering, University of Minnesota, Minneapolis, MN 55455, USA, E-mail: $\{$wtan, chiu0091, parhi$\}$@umn.edu }% <-this % stops a space
\thanks{Antian Wang and Yingjie Lao are with the Holcombe Department of Electrical and Computer Engineering, Clemson University, Clemson, SC 29634, USA. E-mail: $\{$antianw, ylao$\}$@clemson.edu}% <-this % stops a space
}

\begin{document}
%\ninept
%
\date{}
\maketitle

\makeatletter{\renewcommand*{\@makefnmark}{}
\footnotetext{* (equal contribution)}\makeatother}

\begin{abstract}
High-speed long polynomial multiplication is important for applications in homomorphic encryption (HE) and lattice-based cryptosystems. This paper addresses {\em low-latency} hardware architectures for long polynomial modular multiplication using the number-theoretic transform (NTT) and inverse NTT (iNTT). Parallel NTT and iNTT architectures are proposed to reduce the number of clock cycles to process the polynomials. Chinese remainder theorem (CRT) is used to decompose the modulus into multiple smaller moduli. Our proposed architecture, namely PaReNTT,  makes three novel contributions. First, cascaded parallel NTT and iNTT architectures are proposed such that any buffer requirement for permuting the  product of the NTTs before it is input to the iNTT is eliminated. This is achieved by using different {\em folding sets} for the NTTs and iNTT.
Second, a novel approach to expand the set of feasible special moduli is presented where the moduli can be expressed in terms of a few signed power-of-two terms. Third, novel architectures for pre-processing for computing residual polynomials using the CRT and post-processing for combining the residual polynomials are proposed. These architectures significantly reduce the area consumption of the pre-processing and post-processing steps. The proposed long modular polynomial multiplications are ideal for applications that require low latency and high sample rate such as in the cloud, as these feed-forward architectures can be pipelined at arbitrary levels. Pipelining and latency tradeoffs are also investigated. Compared to a prior design, the proposed architecture reduces latency by a factor of 49.2, and the area-time products (ATP) for the lookup table and DSP, ATP(LUT) and ATP(DSP), respectively, by 89.2\% and 92.5\%. Specifically, we show that for n=4096 and a 180-bit coefficient, the proposed 2-parallel architecture requires 6.3 Watts of power while operating at 240 MHz, with 6 moduli, each of length 30 bits, using Xilinx Virtex Ultrascale+ FPGA.
\end{abstract}


\section{Introduction}
Privacy-preserving protocols and the security of the information are essential for cloud computing. To this end, cloud platforms typically encrypt the data by certain conventional symmetric-key or asymmetric-key cryptosystems to protect user privacy. However, these methods cannot prevent information leakage during the computation on the cloud since the data must be decrypted before the computation. To further enhance privacy, homomorphic encryption (HE) has emerged as a promising tool that can guarantee the confidentiality of information in an untrusted cloud. Homomorphic encryption is also deployed in privacy-preserving federated learning~\cite{wibawa2022homomorphic} and neural network inference~\cite{chen2019efficient}.

Homomorphic multiplication and homomorphic addition are two fundamental operations for the HE schemes. Most of the existing HE schemes are constructed from the ring-learning with errors (R-LWE) problem~\cite{lyubashevsky2010ideal} that adds some noise to the ciphertext to ensure post-quantum security. However, the quadratic noise growth of homomorphic multiplication requires the ciphertext modulus to be very large, which results in inefficient arithmetic operations. One possible solution to address this issue is to decompose the modulus and execute it in parallel. This approach has been used in residue number system (RNS) representation. In the literature, RNS-based implementations have been employed in several software~\cite{halevi2014algorithms,chen2017simple} and hardware implementations~\cite{roy2018hepcloud,roy2019fpga,riazi2020heax}. However, RNS relies on the Chinese remainder theorem (CRT), which requires additional pre-processing and post-processing operations. The hardware building blocks for these steps need to be optimized; otherwise, the complexity of the RNS system will negate the advantages of parallelism of the RNS. Meanwhile, modular polynomial multiplication is one of the essential arithmetic operations for the R-LWE problem-based cryptosystems and, indeed, HE schemes. The complexity of the number-theoretic transform (NTT)-based modular polynomial multiplication can be reduced dramatically compared to the schoolbook-based modular polynomial multiplication.

Different modular long polynomial multiplier architectures can be adopted for different applications. For example, a low-area time-multiplexed architecture is well-suited for an edge device. However, the cloud requires very high-speed architectures where multiple coefficients of the polynomial need to be processed in a clock cycle. This inherently requires a parallel architecture where the level of parallelism corresponds to the number of coefficients processed in a clock cycle. While substantial research has been devoted to designing and implementing sequential and time-multiplexed architectures, much less research on parallel NTT-based architectures has been presented. Computing the inverse NTT (iNTT) of the product of NTT of the two polynomials can lead to long latency and extra buffer requirement if its scheduling aspects are not considered as the product needs to be shuffled before the iNTT is computed.

Although parallel NTT-based architectures can achieve low latency and high speed, these require a large silicon area for the arithmetic operations as the word-lengths of the coefficients can be large. To reduce the area, residue arithmetic is used to convert the coefficient into several smaller coefficients that can be implemented using shorter word-lengths. This paper proposes parallel residue arithmetic and NTT-based modular long polynomial multiplication referred to as PaReNTT. The use of different scheduling (folding) of the NTT and iNTT operations eliminates the need for additional buffers. Thus, the latency of the complete operation is reduced. The use of parallel NTT architecture reduces the number of clock cycles needed to process the long polynomial modular multiplication. The proposed parallel NTT and iNTT architectures are completely feed-forward and achieve full hardware utilization. These can be pipelined at any arbitrary level. 
To the best of our knowledge, the proposed architecture is the first approach for a generalized, feed-forward, and parallel NTT-based implementation that eliminates intermediate shuffling or buffer requirement.

The contributions of this paper are three-fold and are summarized below. 
\begin{itemize}[topsep=3pt,itemsep=3pt]
    
    \item Our proposed cascaded NTT-iNTT architecture does not require intermediate shuffling operations. Different folding sets for the NTT and iNTT are used such that the product of the two NTTs can be processed immediately in the iNTT. This leads to a significant reduction in latency and completely eliminates the need for any intermediate buffer requirement. 
    \item We consider special format of primes for the CRT to reduce the cost of the implementation. Specifically, all the primes are not only NTT-compatible but also CRT-friendly and have low Hamming weights (i.e., these contain only a few signed power-of-two terms). Traditional selection of moduli to satisfy these constraints can limit the number of moduli available. A novel approach is proposed to expand the set of moduli that satisfy these constraints. This enables HE architectures for long word-length coefficients.
    \item Novel optimized architectures for pre-processing and post-processing for residue arithmetic are proposed; these architectures reduce area and power consumption. Finally, the low-cost pre-processing and post-processing blocks for the residue arithmetic are integrated into the parallel NTT-based modular polynomial multiplier to achieve high speed, low latency, and low area designs. 

\end{itemize}

The rest of this paper is organized as follows: \secref{math} reviews the mathematical background for HE, RNS, and NTT-based polynomial modular multiplication and the corresponding hardware architectures in prior works. \secref{sec_ntt} presents a parallel architecture for the NTT-based polynomial multiplication that eliminates intermediate storage requirements. Then, \secref{sec_crt} introduces our optimized RNS and CRT-based polynomial multiplier. The performance of our proposed architecture is presented and analyzed in \secref{result}. Finally, \secref{conclusion} concludes the paper. 
\section{Background}\seclbl{math}
\subsection{Notation}
For a polynomial ring $R_{n,q} = \mathbb{Z}_q[x]/(x^n+1)$, its coefficients have to be modulo $q$ (i.e., these lie in the range $[0,q-1]$) and the degree of the polynomial is less than $n$ ($n$ is a power-of-two integer). To ensure all the intermediate results belong to the polynomial ring, a modular reduction operation is needed, which is expressed as ``mod $(x^n+1,q)$'' or $[\cdot]_{q}$.
The polynomial of the ring $R_{n,q}$ is denoted as $a(x) = \sum_{j=0}^{n-1} a_j x^j$, where the $j$-th coefficient of the polynomial $a(x)$ is represented as $a_j$. 

The addition and multiplication of two polynomials modulo $(x^n+1,q)$ (i.e., modular polynomial addition and multiplication) are written as $a(x)+  b(x) $ and $a(x) \cdot b(x) $, respectively. We also use $\odot$ to denote the point-wise multiplication over $(x^n+1,q)$ between two polynomials. Parameters $m = \log_2n$ and $s \in[0,m-1]$ represent the total number of stages and the current stage in the NTT (iNTT), respectively. 

\subsection{Homomorphic encryption}
HE allows the computations (e.g., multiplication, addition) directly on the ciphertext, without decryption, so that the users can upload their data to any (even untrusted) cloud servers while preserving privacy. The HE schemes can be broadly classified as fully HE (FHE) and somewhat HE (SHE). The FHE schemes allow an arbitrary number of homomorphic evaluations while suffering from high computational complexity~\cite{gentry2009fully}. SHE is an alternative solution with better efficiency than the FHE, which only allows performing a limited number of operations without decryption~\cite{lyubashevsky2010ideal,fan2012somewhat,cheon2017homomorphic}.

High-level steps for HE schemes can be summarized in four stages: key generation, encryption, evaluation, and decryption. In particular, the key generation step is used to output three keys: the secret key, public key, and relinearization key, based on the security parameter $\lambda$. Then, using the public key, the encryption algorithm encrypts a message into a ciphertext $ct$. During the evaluation step, a secure evaluation function performs a computation homomorphically for all input ciphertexts and outputs a new ciphertext $ct'$ using the relinearization key. Finally, the result can be obtained using the secret key and $ct'$ in the decryption step. 

Key generation, encryption, and decryption steps are generally executed by the client. Meanwhile, the evaluation step is distributed to the cloud server for homomorphic computation. Different homomorphic evaluation functions have different computational costs. The homomorphic addition is relatively simple since it is implemented by modular polynomial additions. However, homomorphic multiplication requires expensive modular polynomial multiplication. Thus the hardware or software accelerations for the modular polynomial multiplier, especially under the HE parameters with large degrees of polynomial and long word-length coefficients, are demanding. 
\subsection{Residue number system}
To implement homomorphic encryption in various applications, the depth of homomorphic multiplication is one of the main factors that needs to be investigated, and increases proportionally with respect to the word-length of the coefficient. As an example, performing a depth of four homomorphic multiplications with an 80-bit security level requires a 180-bit ciphertext modulus and length-4096 polynomial in prior works~\cite{roy2019fpga}. However, the computation involving the long word-length coefficients is not trivial, which is also inefficient without high-level transformations. Since the moduli in most widely-used SHE schemes, e.g., BGV~\cite{lyubashevsky2010ideal}, BFV~\cite{fan2012somewhat}, CKKS~\cite{cheon2017homomorphic}, are not restricted to be primes, it is possible to choose each modulus to be a product of several distinct primes by using CRT, where each prime is an NTT-compatible prime with a small word-length.  

The CRT algorithm decomposes $q$ to $q_1,q_2,\hdots,q_t$ (i.e., $q = \prod_{i=1}^t q_i $, $q_i$'s are mutually co-prime), and the ring isomorphism $R_q \equiv R_{q_1} \times R_{q_2},\hdots, \times R_{q_t}$. After this decomposition, ring operation in each $R_{q_i}$ is performed separately, which thus can be executed in parallel. From the implementation perspective, the larger the parameter $t$, the smaller each $q_i$ and the simpler arithmetic operation over $R_{q_i}$. 

\subsection{NTT-based polynomial multiplication}
In addition to the long word-length of the coefficient, the long polynomial degree $n$ can be in the range of thousands for the HE schemes to maintain the high security level, which becomes the bottleneck for the implementations in both software and hardware~\cite{aysu2013low,dai2015cuhe}. Therefore, an efficient NTT-based polynomial multiplication method with the time complexity of $\mathcal{O}(n\log n)$ is used. 

To compute $p(x) = a(x) \cdot b(x) \mod (x^n+1,q)$, polynomials $a(x)$ and $b(x)$ are first mapped to their NTT-domain polynomials $A(x)$ and $B(x)$. 
For instance, the NTT computation for polynomial $a(x)$ is expressed as: $A_k = \sum_{j=0}^{n-1} a_j\omega_{n}^{kj}\mod q$, where $k \in [0,n-1]$. $\omega$ is the primitive $n$-th root of unity modulo $q$ (i.e., twiddle factor), which satisfies $\omega^{n} \equiv 1 \mod q$. 
Subsequently, an efficient point-wise multiplication between $A(x)$ and $B(x)$, yields $P(x) = A(x) \odot B(x)$. 
The final result is obtained via iNTT computation: ${p}(x) = n^{-1}\sum_{j=0}^{n-1} P_j\omega_{n}^{-kj}\mod q$, where $k \in [0,n-1]$.

This method significantly reduces the time complexity compared to the $\mathcal{O}(n^{2})$ complexity method of the schoolbook polynomial multiplication along with the modular polynomial reduction. However, this original method involves zero padding of length $n$, which doubles the length of the polynomial in the NTT/iNTT computation. It has been shown that using {\em negative wrapped convolution (NWC)} zero padding can be completely eliminated~\cite{lyubashevsky2008swifft}. However, this requires that the inputs and outputs need to be weighted. These additional weight operations can be eliminated by reformulation of the algorithm. This is referred to as {\em low-complexity negative wrapped convolution} ~\cite{zhang2020highly,longa2016speeding}. A review of the original and low-complexity NTT approaches is presented in \cite{chiu2023nttbased}. Flow graphs for low-complexity NWC based NTT and iNTT are shown in  \figref{ntt_eight} for $n=8$.
\begin{figure}[htbp]
\centering
\resizebox{0.37\textwidth}{!}{
\includegraphics{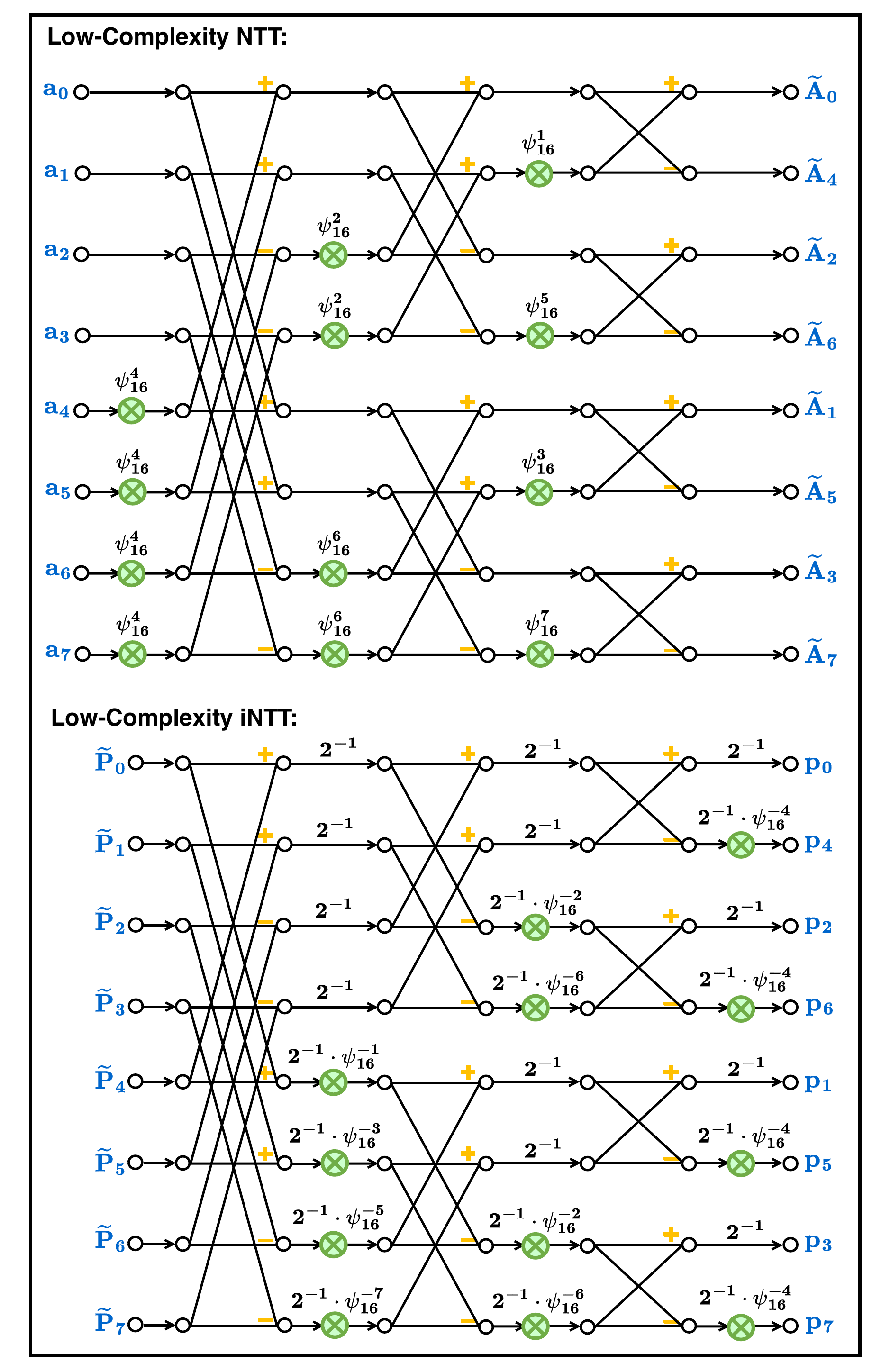}}
\caption{Data-flow graphs for low-complexity NTT and iNTT for polynomial multiplication when $n=8$.}
\figlbl{ntt_eight}
\vspace{-1.5em}
\end{figure}
In particular, $\psi_{2n}$ is the primitive $2n$-th root of unity modulo $q$, and $q$ must be NTT-compatible prime that satisfy that $(q-1)$ is divisible by $2n$. This low-complexity algorithm reduces the number of modular integer multiplications. Meanwhile, the modular multiplication by $2^{-1}$ can be efficiently implemented using low-cost modular adders and a multiplexer (MUX) in hardware (see supplementary information).

\subsection{Prior hardware implementations}\label{priorwork}
In the literature, several hardware architectures based on CRT-based optimization have been proposed in~\cite{roy2019fpga,roy2018hepcloud,xin2021multi,turan2020heaws}, where these architectures are based on feedback architectures with loops for executing multiple operations in a time-multiplexed manner~\cite{roy2019fpga,roy2018hepcloud,xin2021multi}.

The works in~\cite{roy2019fpga,roy2018hepcloud} introduce an approximate CRT method for the BFV scheme, which involves the lifting and scaling operations to switch between a small ciphertext modulus $q$ and a large ciphertext modulus $Q$. Later, a multi-level parallel accelerator utilizing the RNS and NTT algorithms is presented in~\cite{xin2021multi}. Nevertheless, these works mainly focus on optimizing the NTT blocks but not on the CRT system's pre-processing and post-processing functional blocks. 

In particular, the prior works consider a unified architecture for the NTT and iNTT architecture, which is typically constructed from a memory-based or folded architecture framework~\cite{xin2021multi,xing2021compact,zhang2020highly,mert2022medha,mert2020fpga,paludo2022ntt}. This design strategy can reduce the number of required processing elements (PEs). Different from the prior works, our proposed architecture exploits a feed-forward and parallel architecture. Our proposed architecture has no feedback loops/data paths. Therefore, the intermediate results can be executed and passed through to the next stage PE directly. It may be noted that the design of parallel-pipelined  and  memory-based NTT/iNTT architectures are similar to those for prior FFT/iFFT designs~\cite{ayinala2011pipelined,ayinala2013place}.

\section{Parallel NTT-based polynomial multiplier without shuffling operations}\seclbl{sec_ntt}

In this section, we propose a novel real-time, feed-forward, high-throughput, and parallel NTT-based polynomial multiplication architecture design that does not require intermediate shuffling, as shown in \figref{polymult}. For a conventional implementation, an additional shuffling circuit is typically used for reordering output data before computing iNTT. However, such a shuffling circuit requires a large number of clock cycles and registers. In this proposed novel architecture, the two-parallel products are fed into a two-parallel iNTT architecture such that no intermediate buffer is needed. Thus, the outputs of the product are executed immediately by the iNTT. This is possible as we can select different folding sets for the NTT and iNTT. 
Moreover, the proposed architecture is generalized for any value of $n$, parameterized, and can achieve an arbitrary level of pipelining to achieve high-speed operation. 
\begin{figure}[htbp]
\centering
\resizebox{0.35\textwidth}{!}{
\includegraphics{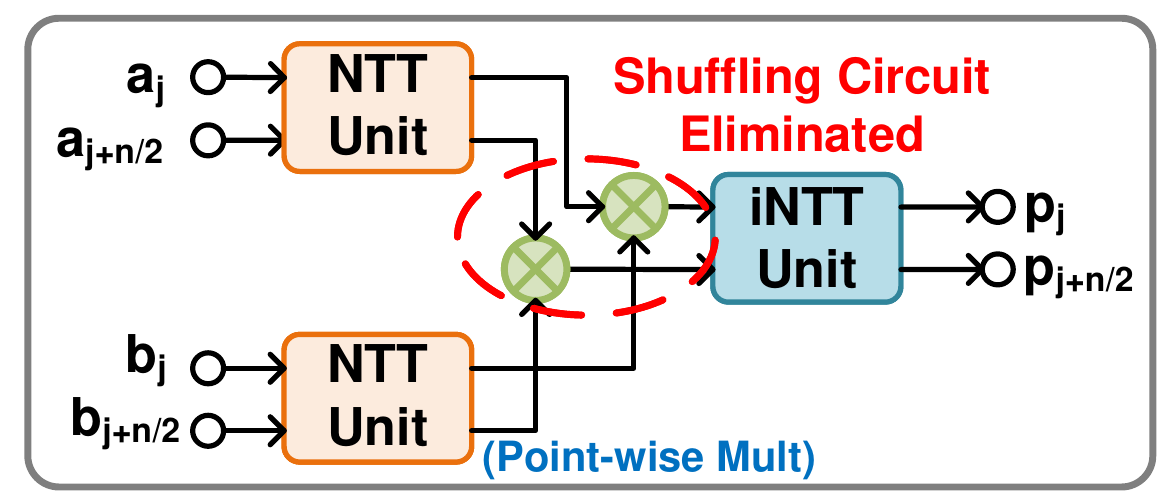}}
\caption{Architecture for the modular polynomial multiplier using two-parallel NTT and iNTT.}
\figlbl{polymult}
\vspace{-1em}
\end{figure}

In particular, the NTT/iNTT units in \figref{polymult} are based on a two-parallel architecture; these are derived using appropriate folding sets and the folding transformation~\cite{parhi2007vlsi,parhi1992synthesis}. \figref{dfg_ntt} and \figref{dfg_intt} show the data-flow graphs for 16-point forward NTT of $a(x)$ and iNTT for $P(x)$, respectively, where each circle represents one butterfly operation. 

After applying the folding transformation, the operations in the same color, i.e., in the same stage, are processed by the same PE and then executed in a time-multiplexed manner. The order in which the butterfly operations are executed in the same PE is referred to as the folding order. Also, the corresponding clock cycle for each butterfly operation is highlighted in blue in \figref{dfg_ntt} and \figref{dfg_intt}. In this 16-point example, the folding set (i.e., the ordered set of operations executed in each PE) of the forward NTT is expressed as:
\begin{alignat}{6}
\mathcal{A}&=\{\mathcal{A}_0,&\mathcal{A}_1,&\mathcal{A}_2,&\mathcal{A}_3,&\mathcal{A}_4,&\mathcal{A}_5,&\mathcal{A}_6,&\mathcal{A}_7&\}\nonumber\\
\mathcal{B}&=\{\mathcal{B}_4,&\mathcal{B}_5,&\mathcal{B}_6,&\mathcal{B}_7,&\mathcal{B}_0,&\mathcal{B}_1,&\mathcal{B}_2,&\mathcal{B}_3&\}\nonumber\\
\mathcal{C}&=\{\mathcal{C}_2,&\mathcal{C}_3,&\mathcal{C}_4,&\mathcal{C}_5,&\mathcal{C}_6,&\mathcal{C}_7,&\mathcal{C}_0,&\mathcal{C}_1&\}\nonumber\\
\mathcal{D}&=\{\mathcal{D}_1,&\mathcal{D}_2,&\mathcal{D}_3,&\mathcal{D}_4,&\mathcal{D}_5,&\mathcal{D}_6,&\mathcal{D}_7,&\mathcal{D}_0&\}.
\eqnlbl{eq_ntt_fold}
\end{alignat}
\begin{figure}[htbp]
\centering
\resizebox{0.45\textwidth}{!}{
\includegraphics{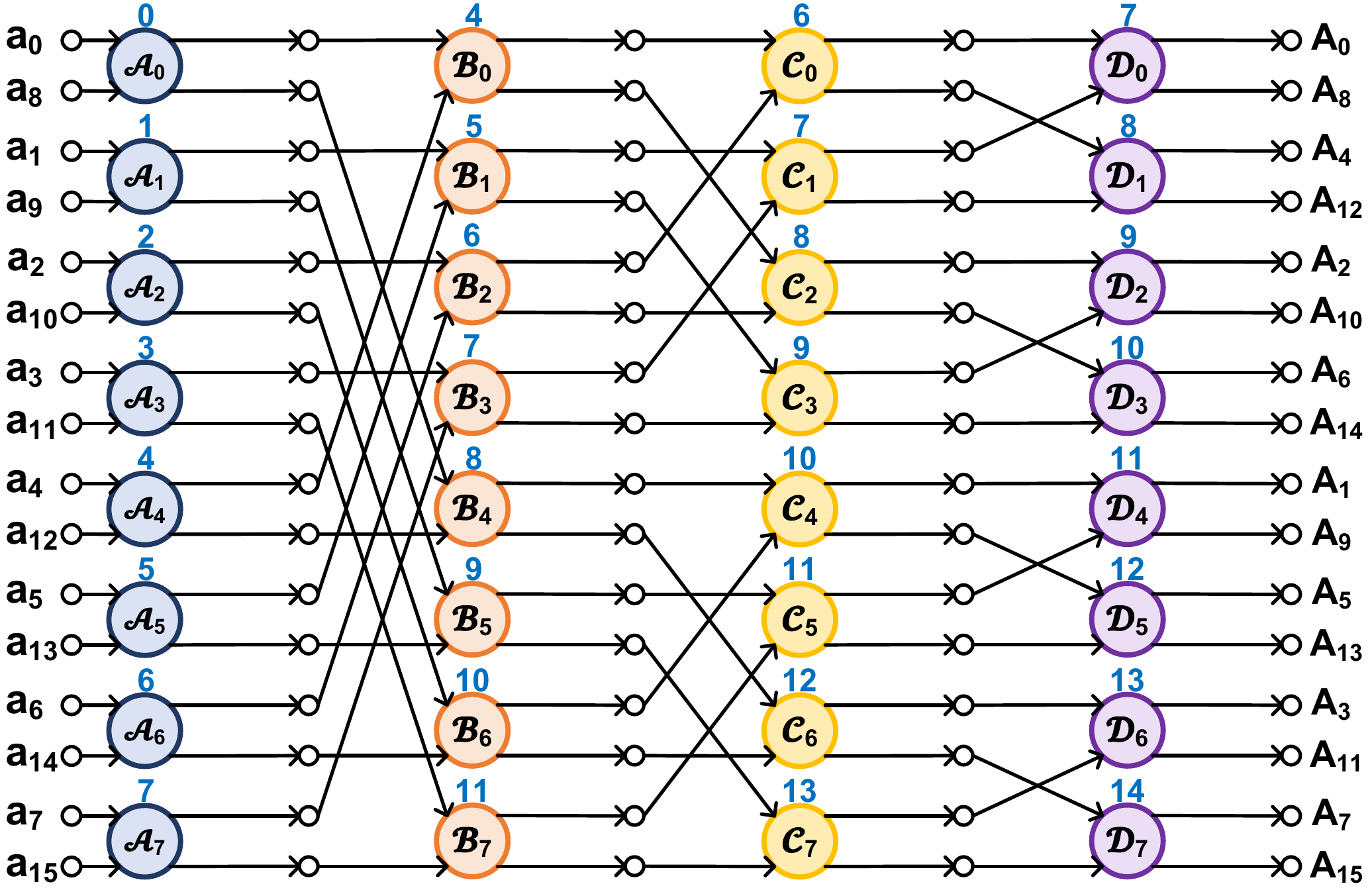}}
\caption{Data-flow graph of the 16-point forward NTT.}
\figlbl{dfg_ntt}
\vspace{-0.5em}
\end{figure}

\begin{figure}[htbp]
\centering
\resizebox{0.45\textwidth}{!}{
\includegraphics{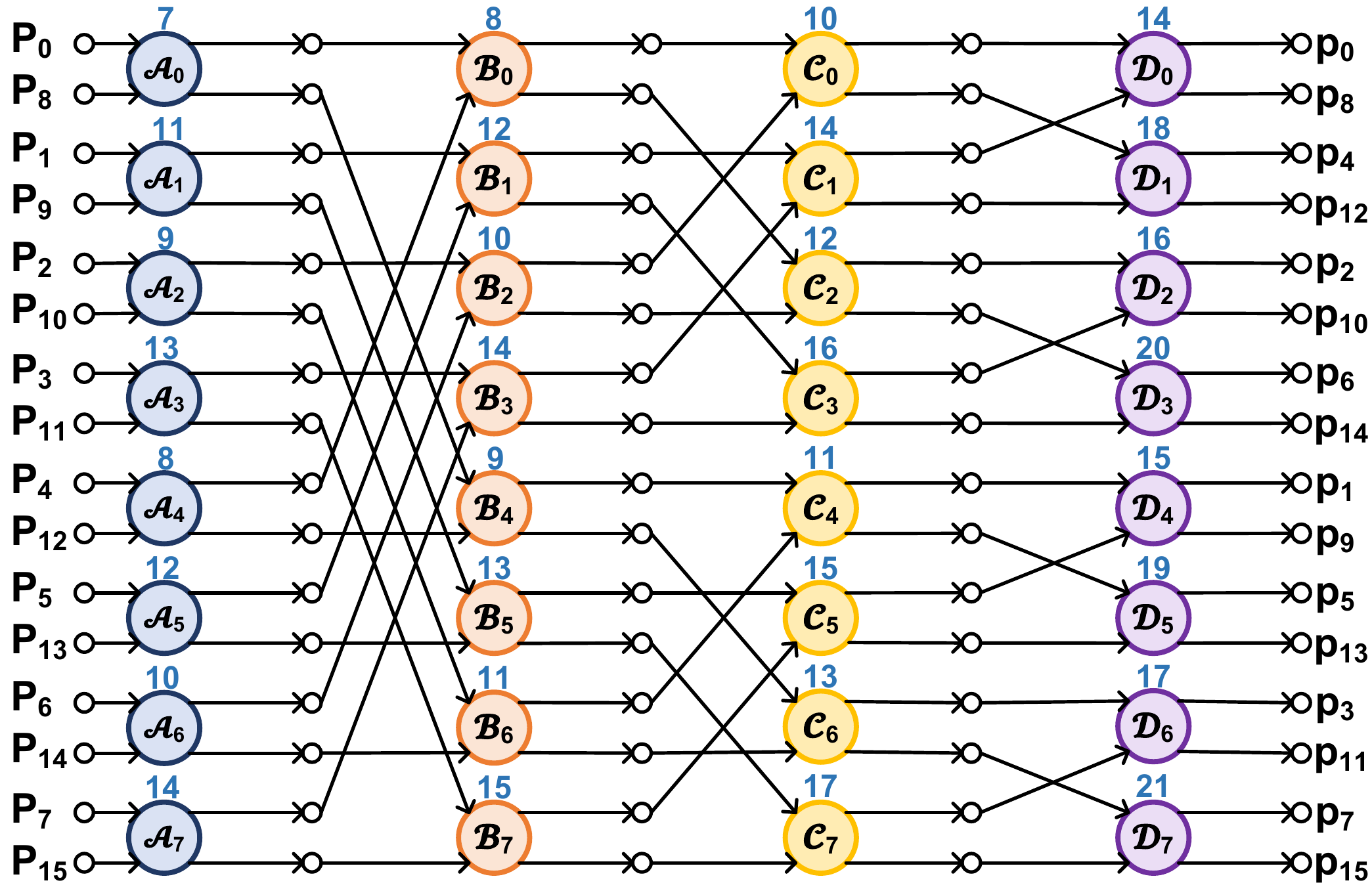}}
\caption{Data-flow graph of the 16-point iNTT. }
\vspace{-1em}
\figlbl{dfg_intt}
\end{figure}

\begin{figure*}[htbp]
\centering
\resizebox{0.7\textwidth}{!}{
\includegraphics{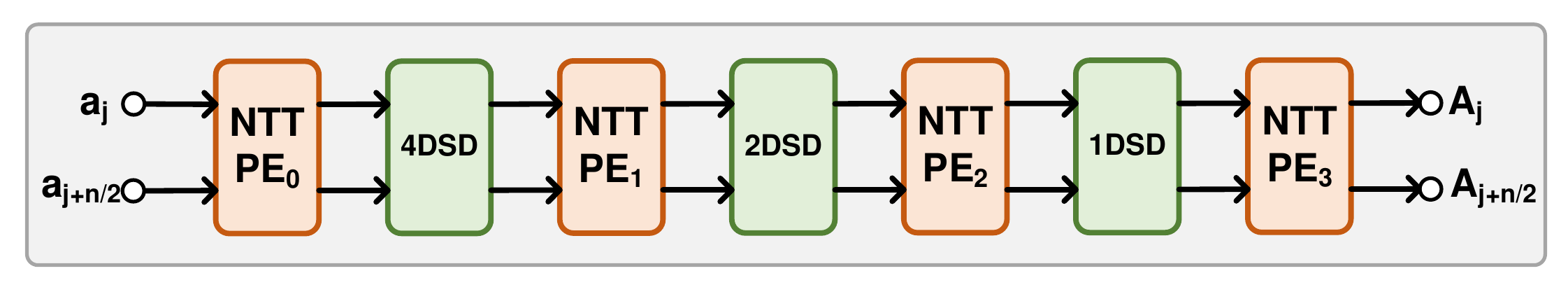}}
\caption{Architecture of the 16-point forward NTT unit.}
\figlbl{arch_ntt}
\vspace{-1.5em}
\end{figure*}
 \begin{figure}[htbp]
\centering
\resizebox{0.40\textwidth}{!}{
\includegraphics{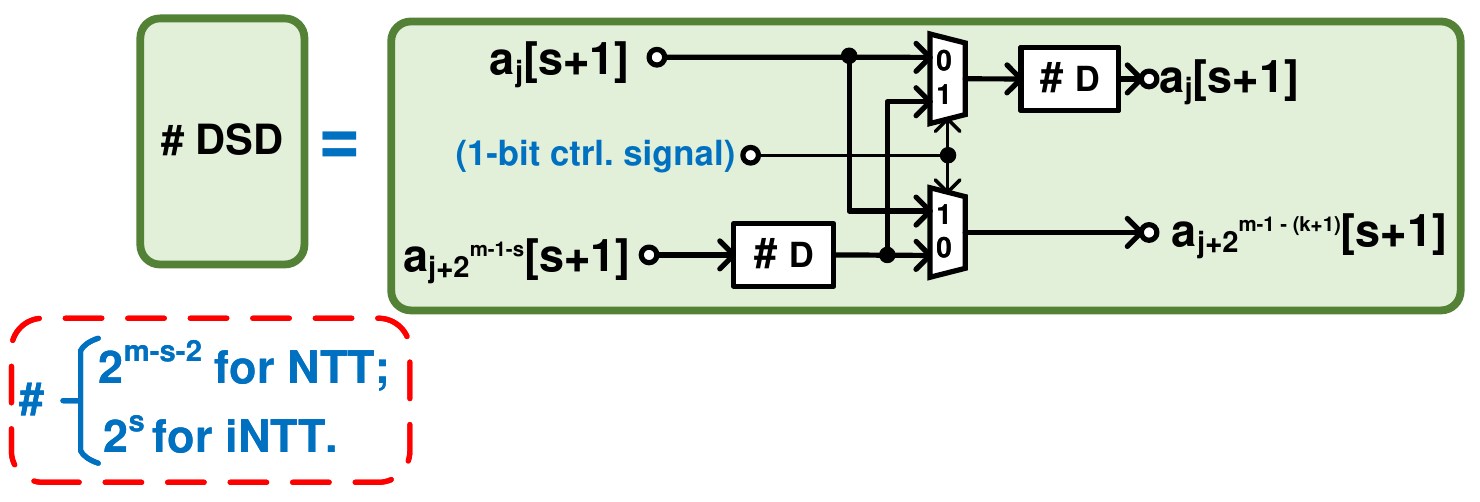}
}
\caption{Architecture for delay-switch-delay (DSD) unit.}
\figlbl{dsd}
\vspace{-1.0em}
\end{figure}
In order to avoid intermediate buffer or data format conversion from NTT to iNTT, the output samples from the last PE in the NTT unit should be fed into the first PE in the iNTT unit at the same clock cycle. This is achieved using the following folding set for the iNTT:
\begin{alignat}{6}
\mathcal{A}&=\{\mathcal{A}_4,&\mathcal{A}_2,&\mathcal{A}_6,&\mathcal{A}_1,&\mathcal{A}_5,&\mathcal{A}_3,&\mathcal{A}_7,&\mathcal{A}_0\}\nonumber\\
\mathcal{B}&=\{\mathcal{B}_0,&\mathcal{B}_4,&\mathcal{B}_2,&\mathcal{B}_6,&\mathcal{B}_1,&\mathcal{B}_5,&\mathcal{B}_3,&\mathcal{B}_7\}\nonumber\\
\mathcal{C}&=\{\mathcal{C}_3,&\mathcal{C}_7,&\mathcal{C}_0,&\mathcal{C}_4,&\mathcal{C}_2,&\mathcal{C}_6,&\mathcal{C}_1,&\mathcal{C}_5\}\nonumber\\
\mathcal{D}&=\{\mathcal{D}_2,&\mathcal{D}_6,&\mathcal{D}_1,&\mathcal{D}_5,&\mathcal{D}_3,&\mathcal{D}_7,&\mathcal{D}_0,&\mathcal{D}_4\}.
    \eqnlbl{eq_intt_fold}
\end{alignat}
\begin{figure*}[htbp]
\centering
\resizebox{0.7\textwidth}{!}{
\includegraphics{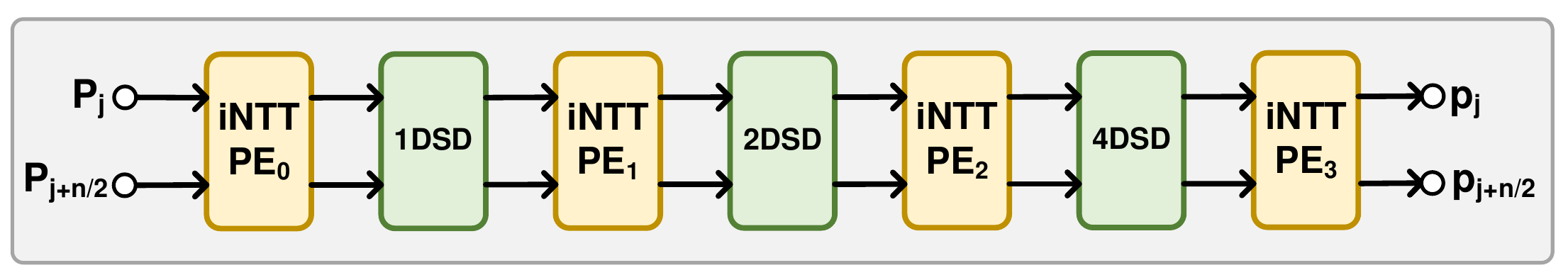}}
\caption{Architecture of the 16-point iNTT unit.}
\figlbl{arch_intt}
\vspace{-1.5em}
\end{figure*}

The NTT and iNTT designs are inspired by the design of parallel FFT architectures based on folding sets~\cite{ayinala2011pipelined}. Parallel NTT architectures based on folding sets was presented in our earlier work~\cite{tan2021pipelined}. The NTT architecture in~\figref{arch_ntt} is derived using the folding sets shown in~\eqnref{eq_ntt_fold}. Specifically, this architecture has four PEs and three delay-switch-delays (DSDs), where the structures for PE and DSD are illustrated in~\figref{ntt_bf} and~\figref{dsd}. 
Besides, the DSD block utilizes two MUXs and two register sets, such that it can store the specific data in the data-path and then either switch or pass the data to the PE. Note that the number of registers inside each register set is varied in different stages. In the $s$-th stage, each register set has $2^{m-s-2}$ registers in the DSD block for the NTT architecture. 
\begin{figure}[htbp]
\centering
\resizebox{0.40\textwidth}{!}{
\includegraphics{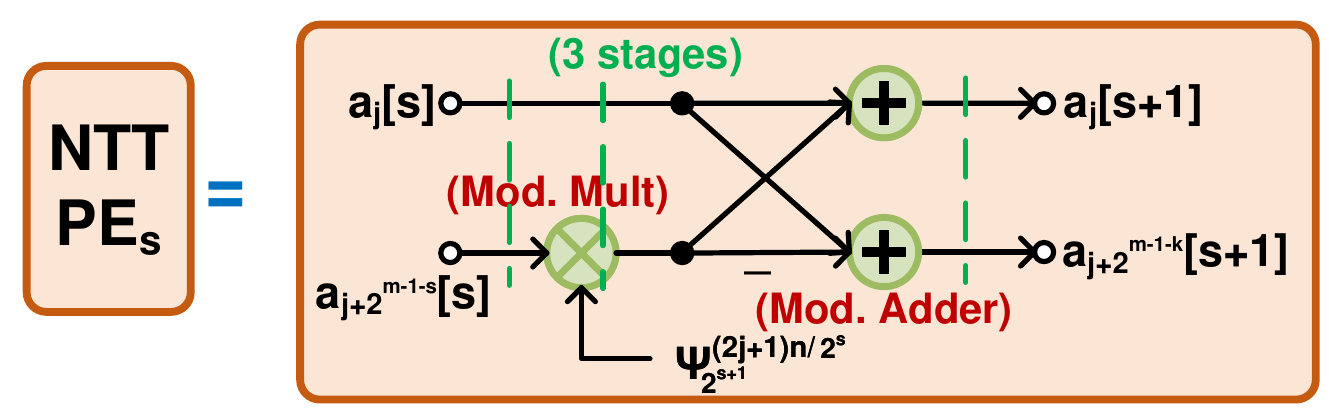}}
\caption{Architecture for DIT-based butterfly with merging the weighted operation in NTT. Pipelining cut sets marked in green lines. }
\figlbl{ntt_bf}
\end{figure}
Furthermore, the architecture for iNTT is shown in \figref{arch_intt}, and its components are described in \figref{intt_bf} and \figref{dsd}. \figref{intt_bf} shows a hardware-friendly PE design for iNTT that only involves one right shift operation, one modular addition with constant $\frac{q+1}{2}$, and one MUX for one modular division by two~\cite{zhang2020efficient}. One of the main differences between NTT and iNTT architectures is the number of registers located inside each DSD block since they are determined by the folding set as in \eqnref{eq_intt_fold}. Specifically, $2^{s}$ registers are required for each register set in the $s$-th stage for the iNTT architecture. Even though the operations of NTT and iNTT are very similar, we consider two separate architectures instead of considering a unified and reconfigurable architecture. The rationale is as follows. Since modular multiplications are heavily used in homomorphic multiplication, using two different architectures for NTT and iNTT allows a continuous flow of the input polynomials and thus can highly accelerate the HE multiplication. 

\begin{figure}[htbp]
\centering
\resizebox{0.40\textwidth}{!}{
\includegraphics{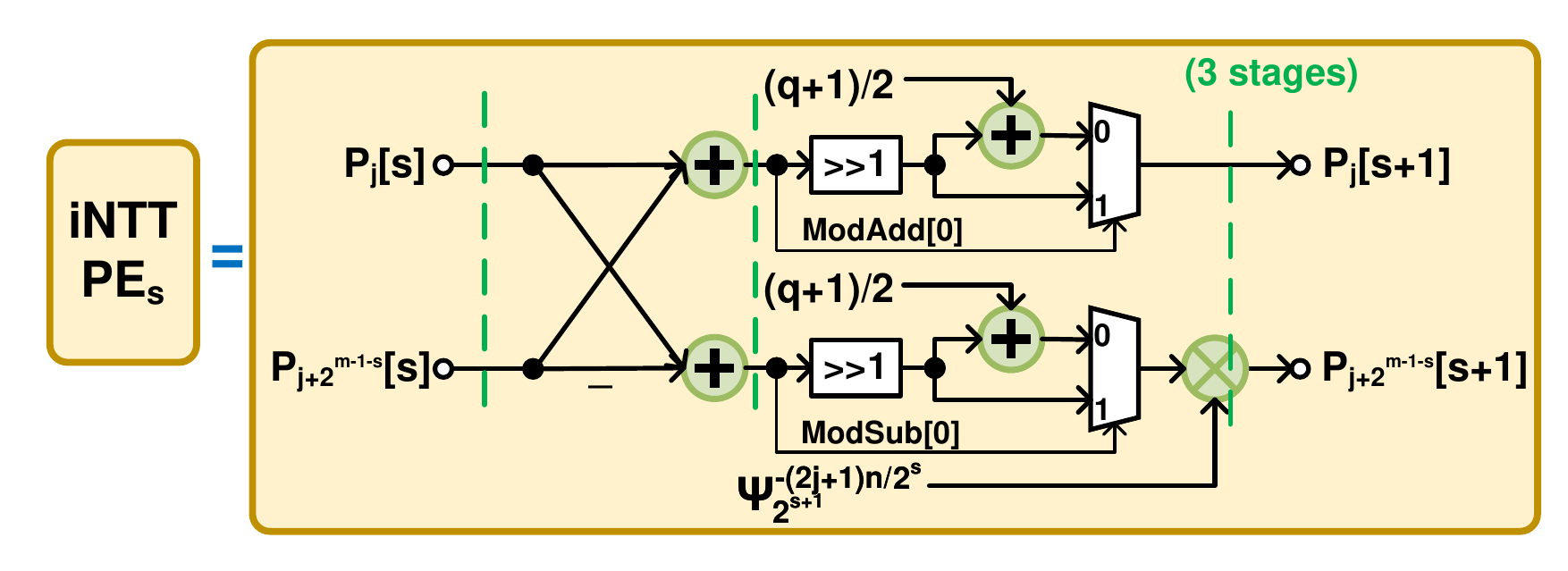}}
\caption{Architecture for DIF-based butterfly with merging the weighted operation in iNTT. Pipelining cut sets marked in green lines. }
\vspace{-1em}
\figlbl{intt_bf}
\end{figure}

The 16-point architectures in \figref{arch_ntt} and \figref{arch_intt} can also be easily generalized to any power-of-two length $n$ by having $m$ PEs and $(m-1)$ DSDs blocks.  Furthermore, the general case NTT and iNTT folding sets are defined as follows. We denote the PE in $s$-th stage as $PE_s$, and the NTT folding set for the butterfly operations performed inside this PE are illustrated in Table \ref{tb_ntt}. The entries in the Table describe the node index of the node of that stage in the data-flow graph. The folding order describes the clock partition at which the node is executed. For example, a folding order $s$ implies that the node is executed at clock cycle $(n/2)l+s$ where $l$ is an integer. The cardinality of the folding set is $n/2$ as there are $n/2$ operations (nodes) in an NTT stage. Thus the scheduling period is $n/2$.

\begin{table*}[htbp]
  \centering
  \caption{Generalized folding order for NTT}\label{tb_ntt}
  \resizebox{0.6\textwidth}{!}{
\begin{tabular}{|c|c c c c c  c|}
\hline
Folding Order &0 & 1  & &$l$ &  &$\frac{n}{2}-1$\\ \hline
$PE_0$ &0 &1 &...& $l$ &... &$\frac{n}{2}-1$ \\ \hline
$PE_1$  &$2^{m-2}$ &$2^{m-2}+1$ &... &$2^{m-2}+ l \mod \frac{n}{2}$ &...  &$2^{m-2}-1 $    \\ \hline
  & & & & ...  & &  \\ \hline
  $PE_s$ &$2^{m-s-1}$ &$2^{m-s-1}+1$ &...  &$2^{m-s-1}+l \mod \frac{n}{2}$ &...  &$2^{m-s-1}-1 \mod \frac{n}{2}$ \\ \hline
  & & & & ...  & &  \\ \hline
  $PE_{m-1}$ &1 &$2$ &... &$l+1$ &...  &$0$ \\ \hline
\end{tabular}
}
\vspace{-1em}
\end{table*}

The folding set for iNTT can also be generalized as in Table \ref{tb_intt}, where the symbol ``$\langle  \cdot \rangle$'' means the bit-reverse representation for the folding set with respect to a $(m-1)$-bit integer (e.g., $\langle 1 \rangle=\langle 001_b \rangle=100_b=4$ when $m=4$). Specifically, if a node $i$ in the NTT has folding order $i$, the folding order of the corresponding node in iNTT is $\langle i \rangle~-~1$ modulo ($n/2)$. While the bit-reversed scheduling has been known to eliminate latency and buffer requirements at the data-flow graph level, the observation that the same property holds in a parallel NTT-iNTT cascade is non-intuitive and new.

Note that if the iNTT was designed using the same folding set in \eqnref{eq_ntt_fold}, the product would need to be input to a DSD of size 4 ($n/4$ in general). This would introduce an additional latency of 4 ($n/4$ in general) clock cycles. The use of different folding sets for NTT and iNTT eliminates any additional DSD circuit and its associated latency.

\begin{table*}[htbp]
  \centering
  \caption{Generalized folding order for iNTT}\label{tb_intt}
\resizebox{0.7\textwidth}{!}{
\begin{tabular}{|c|c c c c c  c|}
\hline
Folding Order &0 & 1  & &$l$ &  &$\frac{n}{2}-1$\\ \hline
$PE_0$ &$\langle 1 \rangle$ &$\langle 2 \rangle$ &...& $\langle l+1 \rangle$ &... &$\langle 0 \rangle$ \\ \hline
$PE_1$  &$\langle 0 \rangle$ &$\langle 1 \rangle$ &... &$\langle l\rangle$ &...  &$\langle 2^{m-1}-1\rangle $    \\ \hline
  & & & & ... &  &  \\ \hline
  $PE_s$ &$\langle 2- 2^s\mod \frac{n}{2} \rangle$ &$\langle 2- 2^s+1 \mod \frac{n}{2} \rangle$ &...  &$\langle 2- 2^s +l \mod \frac{n}{2} \rangle$ &...  &$\langle 2- 2^s -1 \mod \frac{n}{2}\rangle $ \\ \hline
  & & & & ...  & &  \\ \hline
  $PE_{m-1}$ &$\langle 2 \rangle$ &$\langle 3 \rangle$ &...& $\langle l+2 \mod \frac{n}{2} \rangle$ &... &$\langle 1 \rangle$ \\ \hline
\end{tabular}
}
\end{table*}

\section{Moduli Selection and PaReNTT architecture}\seclbl{sec_crt}

\subsection{Overview of proposed PaReNTT architecture}
\figref{overview} shows the overview of the proposed PaReNTT architecture, which can be divided into three constituent steps. The first step, referred to as residual polynomials computations (pre-processing operation), splits the two input polynomials into several polynomials whose coefficients are small. Rather than employing a single modular polynomial multiplier, several modular polynomial multiplications are executed in parallel in the residual domain. Subsequently, the post-processing operation performs the inverse mapping for the product polynomials to one polynomial using the CRT. The result is the same as directly performing the modular polynomial multiplication for two input polynomials. 
\begin{figure}[htbp]
\centering
\resizebox{0.4\textwidth}{!}{
\includegraphics{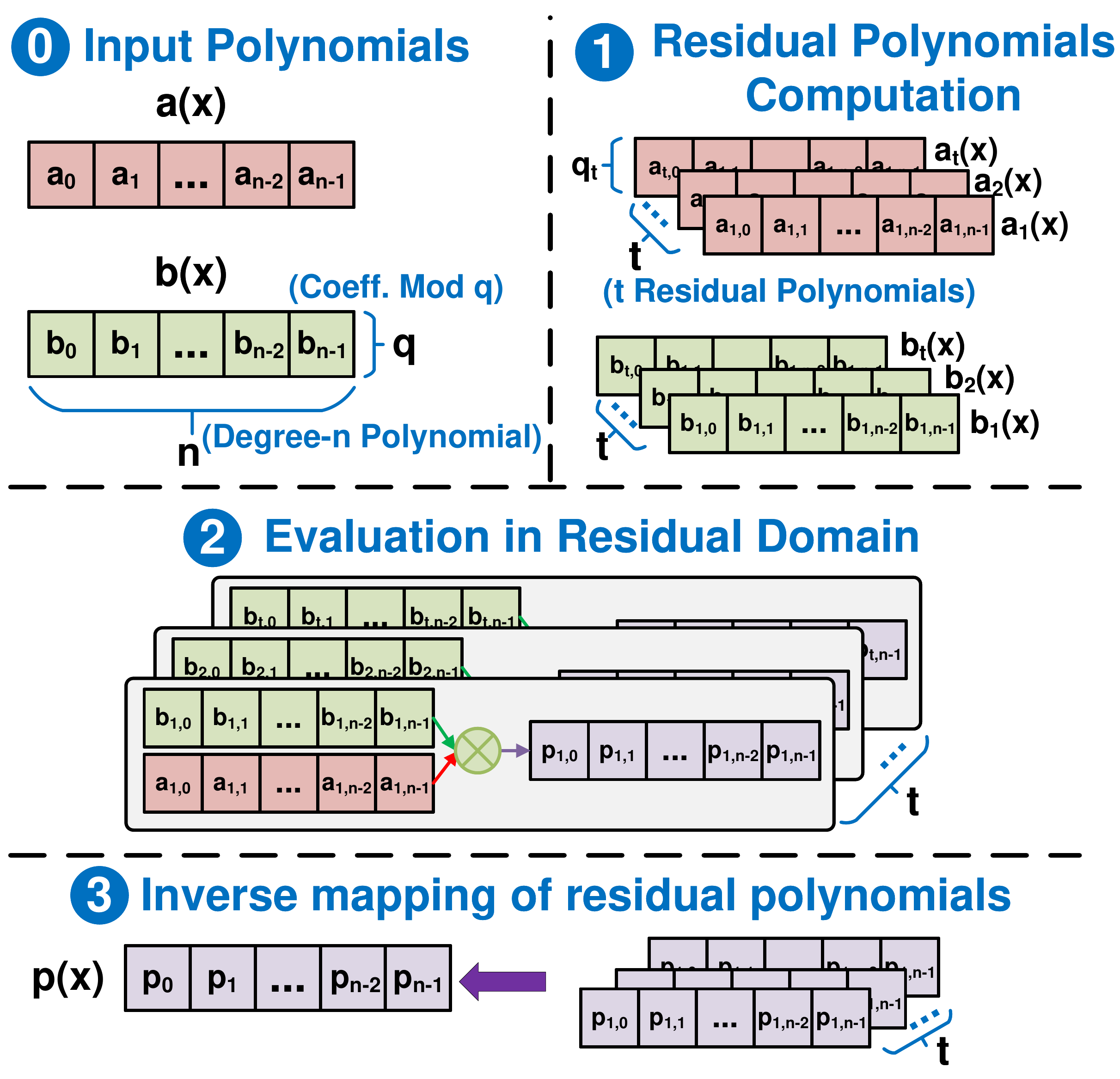}}
\caption{Overview of proposed PaReNTT architecture.}
\figlbl{overview}

\end{figure}

\subsection{Special NTT-compatible and CRT-friendly primes selection}
As opposed to the prior works that randomly select the co-primes, this paper studies and utilizes the property of the special co-primes to reduce the computational cost and the silicon area. The main idea of this optimization is to trade the flexibility of the co-primes selection for the timing/area performance of the architectures. 

In the proposed architecture, each $q_i$ not only needs to be an NTT-compatible prime but also has a short word-length, which is defined as 
\begin{equation}
    q_i = 2^v -\beta_i, \quad \beta_i = 2^{v_{1i}} \pm 2^{v_{2i}}  \pm \hdots \pm 2^{v_{n_q i}}   -1,
    \eqnlbl{eq_phase2}
\end{equation}
where $v$ is the word-length of $q_i$, $v_{1i}>v_{2i}\hdots > v_{n_q i}$. The number of signed power-of-two terms in $q_i$ is $(n_q ~+~2)$.

The special NTT-compatible and CRT-friendly primes can be found through an exhaustive search for the $t$ co-prime factors and then combined to form the $vt$-bit ciphertext modulus, $q$. The two constraints that need to be satisfied are: (1) $(q_i ~-~ 1)$ is a multiple of $2n$ and (2) $\lceil \frac{\mu-1}{n_\beta} \rceil >v_{1i}>v_{2i}$. The second constraint is derived below in Subsection C; see \eqnref{miu_eqn}. Here $\mu$ is the word-length of the
input to the Barrett reduction unit (see description in Subsection C below). In a typical Barrett reduction implementation, $\mu ~=~ 2 v$. In the proposed approach, for given $v$ and $t$, $\mu$ and $n_q$ are increased to expand the number of feasible moduli.

A CRT-friendly modulus leads to an optimized hardware architecture with respect to the overall timing and area performance for the pre-processing and post-processing steps. Our exhaustive search approach generates $q_i$ that are similar to the Solinas prime, and contain a few signed power-of-two terms~\cite{tan2021high,hamburg2015ed448}. 

The integer multipliers have a larger area consumption and longer delay than the integer adders for the hardware implementation. Besides, the area and delay are proportional to the word-length. Therefore, one possible direction to optimize the modular multiplier, pre-processing stage, and post-processing stage architectures is to reduce the number of integer multipliers, especially the long integer multipliers. In the proposed approach, all the integer multipliers are eliminated when multiplying by $q_i$, which significantly reduces the computation cost. 

\vspace{-1em}
\subsection{Residual polynomials computation unit}
The pre-processing stage maps the input polynomials to their residual polynomials by applying the CRT algorithm, as shown in Step 1 of \figref{overview}. For the polynomial $a(x)$, its residual polynomials are 
\begin{equation}
    a_i(x) = [a(x)]_{q_i} =  \sum_{j=0}^{n-1} (a_{i,j}\text{ mod } q_i)x^j,\quad i\in [1,t].
    \eqnlbl{pre-pro}
\end{equation}
A key operation within the pre-processing stage is the execution of modular reduction. One approach to avoiding division operation in computing modular operation is the use of Barrett reduction~\cite{barrett1986implementing}. This is described by:
\begin{align*}
    a \mod q ~=~ a - (\frac{am}{2^\mu}) \cdot q ~= ~a - ((a \epsilon) \gg \mu) \cdot q    
    \label{eq:barrett}
\end{align*}
where $\epsilon = \lfloor \frac{2^\mu}{q} \rfloor$ can be pre-computed, and $\mu$ is the word-length of $a$. 

The prior works employ the \textit{divide-and-conquer paradigm} for residual polynomials computation to enhance the parallelism and reduce the complexity, such as in~\cite{roy2018hepcloud}. An example of this method is shown in \figref{archi-pre-pro}(a), demonstrating a fully parallel implementation for $t=4$. Despite its advantages, this method requires modular multiplication within each segment, presenting opportunities for further optimization. In particular, we exploit the low Hamming weight property of the moduli and replace the modular multipliers by Shift Add Units (SAU).

\begin{figure*}[htbp]
\centering

\resizebox{0.7\textwidth}{!}{
\includegraphics{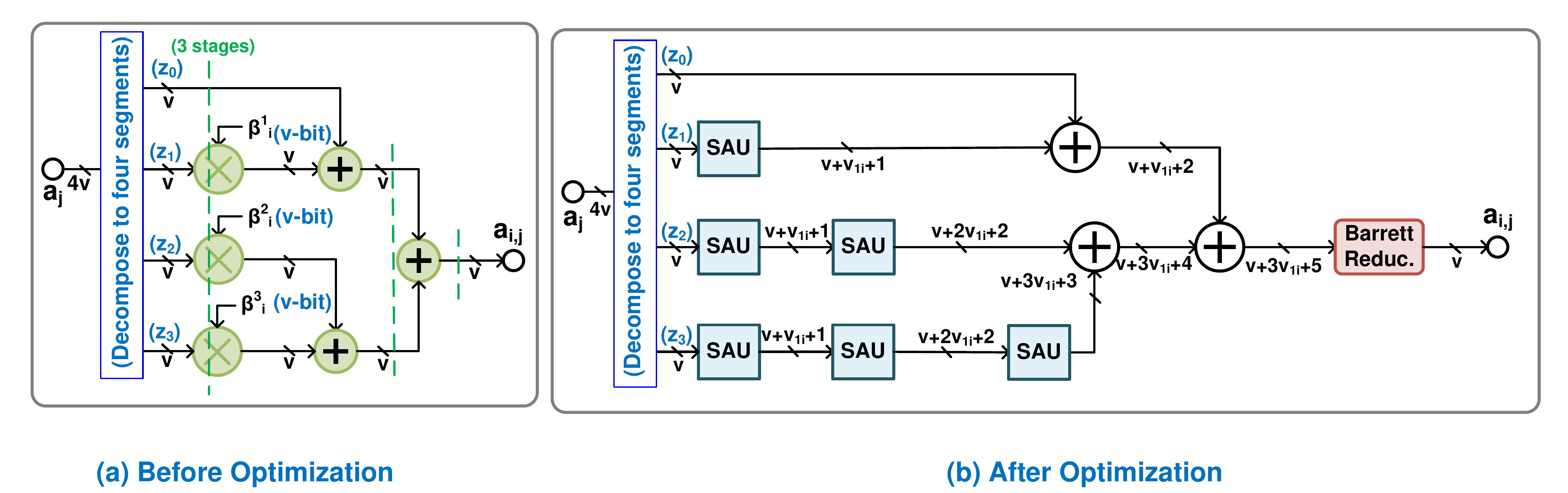}
}
\caption{Top-level architecture of residual coefficient computation unit when $t = 4$. } 
\figlbl{archi-pre-pro}
\end{figure*}
\begin{figure}[htbp]
\centering
\resizebox{0.3\textwidth}{!}{
\includegraphics{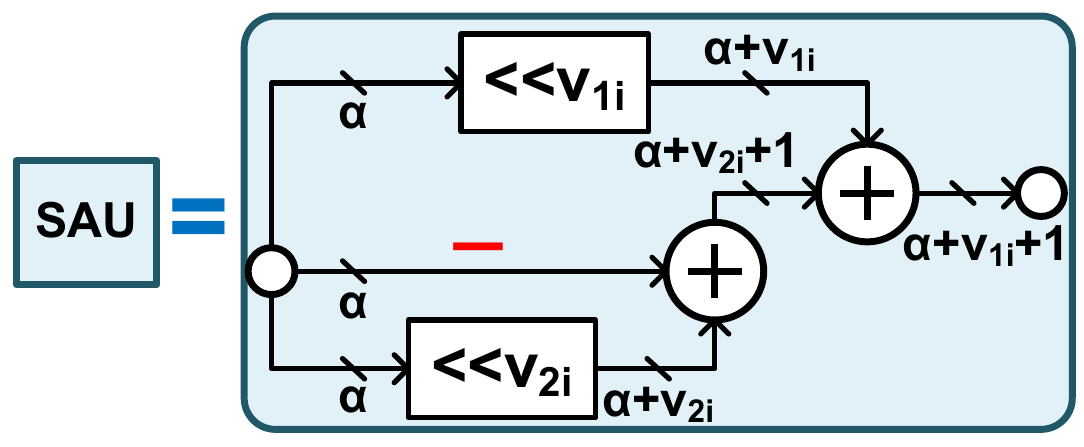}}
\caption{SAU unit of residual coefficient computation unit whose input word-length is $\alpha$.} 
\figlbl{archi-pre-pro_basic_unit}
\vspace{-2em}
\end{figure}
\begin{figure*}[htbp]
\centering
\resizebox{0.99\textwidth}{!}{
\includegraphics{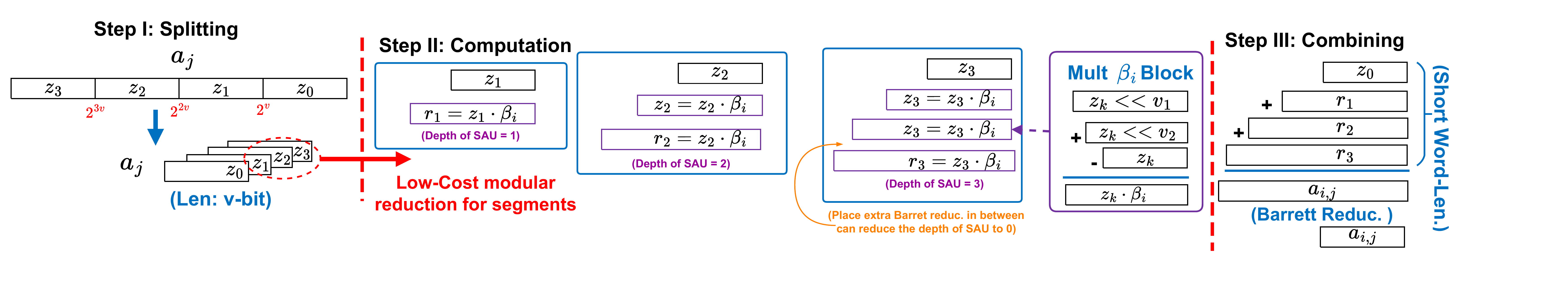}}
\caption{Flow chart for the residual coefficient computation unit when $t = 4$.}
\vspace{-1.5em}
\figlbl{dfg_pre-pro}
\end{figure*}
Algorithm~\ref{alresi} presents our proposed novel and hardware-friendly optimization to implement \eqnref{pre-pro}. The architectures for the prior work and the proposed algorithm are shown in \figref{archi-pre-pro}(a) and \figref{archi-pre-pro}(b), respectively. Similar to the prior work in~\cite{roy2018hepcloud}, Line 1 in Algorithm~\ref{alresi} begins by splitting a large integer $a_{j}$ into several segments where each segment has $v$ bits ($v$ is the word-length of $q_i$). For simplicity, we define the base $B = 2^v$. Thus, each segment within $a_{i,j}$ can be expressed as $z_k \cdot B^k$, $k \in [0,t-1]$. The next step involves the modular reduction for each segment, which is the main focus of our hardware optimization. 

\begin{algorithm}[htbp]
\caption{\textbf{Efficient residual coefficient computation}}
\label{alresi}

\hspace*{\algorithmicindent} \textbf{Input:} $a_j \in [0,q-1]$ and $q_i$

\hspace*{\algorithmicindent} \textbf{Output:} $a_{i,j} = a_j \mod  q_i$, $a_{i,j}\in R_{q_i}$

\begin{algorithmic}[1]
     \STATE $a_j = z_0 + z_1\cdot B + z_2 \cdot B^2+,..., + z_{t-1}\cdot B^{t-1}$ {//$B = 2^v$}
        \FOR{$k = 1$ \TO $t-1$}{
                    \STATE $r_k = z_k\times \beta_i^{k}\quad$ {// $\beta_i = B \mod q_i$} 
        }\ENDFOR
   \STATE $a_{i,j} = z_0 + r_1 + ... +r_{t-1} \mod q_i$ 
\end{algorithmic}
\end{algorithm}

Line 3 in Algorithm~\ref{alresi} no longer requires $v\times v$-bit integer multiplication with $\beta^k_i$ to obtain each $r_k$. Instead, the proposed method utilizes the shift and add operations to eliminate the expensive modular multiplications. 

Besides, different from the baseline design in~\figref{archi-pre-pro}(a) where the modular reductions are required to reduce each $r_k$ modulo $q_i$, our design reduces $(t-1)$ reduction units to only one in the ideal case, as required in Line 4 of Algorithm~\ref{alresi}. The rationale behind this method is as follows. The product $r_k$ in the prior work is a $2v$-bit integer, as $\beta^k_i$ and $z_k$ are both $v$-bits each.

Since $q_i$ only contains a few signed power-of-two terms, a long integer multiplication in Line 3 of Algorithm~\ref{alresi} is replaced by an SAU. For instance, for a special prime $q_i = 2^v - 2^{v_{1i}} - 2^{v_{2i}} +1$, $\beta_i$ in Line 3 of  Algorithm~\ref{alresi} can be expressed as
\begin{equation}
    \beta_i = [2^v]_{q_i} \equiv 2^{v_{1i}} + 2^{v_{2i}} - 1.
    \eqnlbl{beta_i_eq}
\end{equation}
The multiplication by $\beta_i$ using SAU is shown in \figref{archi-pre-pro_basic_unit}.
Here, the word-length of $z_1 \times \beta_i$ is $(v+v_{1i}+1)$. After $n_\beta$ SAUs, the word-length is increased to $(v+n_\beta(v_{1i}+1))$. The word-length of $a_{i,j}$ at Line 4 of Algorithm~\ref{alresi}, $\mu$, is greater than or equal to $(v+n_\beta(v_{1i}+1)+1)$, where $\mu$ is the word-length of the input to the Barrett reduction unit. This leads to the constraint:
\begin{equation}
    \lceil \frac{\mu-1}{n_\beta} \rceil >v_{1i}>v_{2i}.
    \eqnlbl{miu_eqn}
\end{equation}
The parameter $n_\beta=t-1$ in the general case. 

A block diagram to illustrate Algorithm~\ref{alresi} is shown in \figref{dfg_pre-pro}, for $t = 4$. It can be seen  that the modular multiplication in $z_k \times \beta_i^k$ in \figref{archi-pre-pro}(a) can be replaced by the shift and add operations, resulting in reduced hardware costs. Since a multiplier is typically quadratically more expensive than an adder with respect to word-length, using such a shift and add operation is more area efficient than using a multiplier to obtain its result $r_k$.

\subsection{Increasing the number of primes as required}
An increase in the number of co-prime factors $t$ can eventually deepen the depth of the SAU, resulting in a long word-length in the intermediate result, thus yielding inefficient computation. To overcome this bottleneck, two alternative solutions are employed. 

\textbf{Approach 1.} The first solution involves the simple strategic placement of an extra Barrett reduction unit within the data-path, aiming to decrease the maximum depth of SAU. 
Inserting additional Barrett reductions between the SAUs can reduce the depth of SAU to zero and consequently decrease the word-length of the intermediate result to $v$-bit.  For instance, the application of an additional Barrett reduction unit for $r_3$ can minimize the depth of SAU to 1, as shown and highlighted in orange in \figref{crt_pre_more_appro_1} and \figref{dfg_pre-pro}, ensuring all input word-lengths for Barrett reduction units are short. Consequently, as the operating intermediate results $r_k$ are represented using short word-lengths, combining all the $r_k$ and $z_0$ to calculate $a_{i,j}$ requires only adders and a Barrett reduction unit.

Despite this overhead, it maintains a smaller hardware resource requirement than the prior design, as shown in \figref{archi-pre-pro}(a), owing to a reduced number of Barrett reduction units and the elimination of integer multiplication. This method is appropriate when the number of moduli is small (for example, less than 5).

\textbf{Approach 2.} When the number of moduli, $t$, is large, the above approach is not efficient, as the number of SAUs grows in a square manner with $t$. For this case, we propose a novel approach described in Algorithm~\ref{al_more_prime}. First, $t$ is decomposed as $t = d\cdot t'$ where $t'$ moduli are combined using SAUs similar to Approach 1 and form a block. Then $d$ such blocks are used, where each block processes $t'$ moduli. The maximum depth of SAUs in each block is $(t'-1)$.
 
Note that the co-prime factors used in this approach require an adjustment $n_\beta = t'-1$ in \eqnref{beta_i_eq} in order to satisfy the condition $\lceil \frac{\mu-1}{n_\beta} \rceil >v_{1i}>v_{2i}$.

\figref{crt_pre_more} illustrates an example for six co-prime factors ($t=6$). This circuit primarily comprises two blocks ($d=2$) of SAU units (marked in blue), where each block has three inputs ($t'=3$), augmented with additional Barrett reduction units and one multiplier.  In this example, each segment $z_k$ undergoes modular reduction by multiplying $\beta^k_i$, where $k\in[0,5]$. The first block computes the multiplication with $\beta^0_i$ to $\beta^2_i$ by using low-cost SAUs:
\begin{equation}
    sum_{i,0} = z_0\cdot \beta^0_i + z_1\cdot \beta^1_i + z_2\cdot \beta^2_i.
\end{equation}
Since $\beta^0_i=1$, no modular operations are needed for $z_0$. 

Segments $z_3$ to $z_5$ serve to execute the modular reduction, which are subsequently optimized through the application of the distributivity property of multiplication:
\begin{align}
        sum_{i,1} &= [z_3\cdot \beta^3_i + z_4\cdot \beta^4_i + z_5\cdot \beta^5_i]_{q_i}, \nonumber\\
              & = [z_3\cdot \beta^0_i + z_4\cdot \beta^1_i + z_5\cdot \beta^2_i]_{q_i}\cdot [\beta^3_i]_{q_i},
    \eqnlbl{eq_sum1}
\end{align} 
where $[z_3\cdot \beta^0_i + z_4\cdot \beta^1_i + z_5\cdot \beta^2_i]_{q_i}$ can be implemented through identical components in the first block with the SAU units, followed by a Barrett reduction unit, and the multiplication $[\beta^3_i]_{q_i}$ (a $v$-bit pre-computed constant) instantiated through a $(v\times v)$-bit multiplier. 
This novel optimization in \eqnref{eq_sum1} ensures the intermediate result of $sum_{i,1}$ is fixed to $2v$-bit. Ultimately, $sum_{i,0}$ and $sum_{i,1}$ are accumulated and reduced to a $v$-bit result by a Barrett reduction unit. 

\begin{algorithm}[htbp]
\caption{\textbf{Efficient residual coefficient computation by Factorization}}
\label{al_more_prime}

\hspace*{\algorithmicindent} \textbf{Input:} $a_j \in [0,q-1]$, $q_i$, and $t = t'\cdot d$

\hspace*{\algorithmicindent} \textbf{Output:} $a_{i,j} = a_j \mod  q_i$, $a_{i,j}\in R_{q_i}$

\begin{algorithmic}[1]
     \STATE $a_j = z_0 + z_1\cdot B + z_2 \cdot B^2+,..., + z_{t-1}\cdot B^{t-1}$ {//$B = 2^v$}
        \FOR{$ \rho = 0$ \TO $d-1$} { 
        
                    \FOR{$k = 1$ \TO $t'-1$}{                     
                        \STATE $r_k = z_k\times \beta_i^{k}\quad$ {// $\beta_i = B \mod q_i$} 
                    }\ENDFOR
        \IF{$\rho==0$}{
            \STATE $sum_{i,0} = z_0 + r_1 + ... +r_{t'-1}$
        }
        \ELSE {
         \STATE $sum_{i,\rho} = [z_0 + r_1 + ... +r_{t'-1}]_{q_i} \cdot [\beta_i^{t'\rho}]_{q_i}$  
        }  
        \ENDIF
        }\ENDFOR
   \STATE $a_{i,j} = sum_{i,0} + sum_{i,1} + \cdots + sum_{i,d-1} \mod q_i$ 

\end{algorithmic}
\end{algorithm}

\begin{figure}[htbp]
\centering
\resizebox{0.45\textwidth}{!}{
\includegraphics{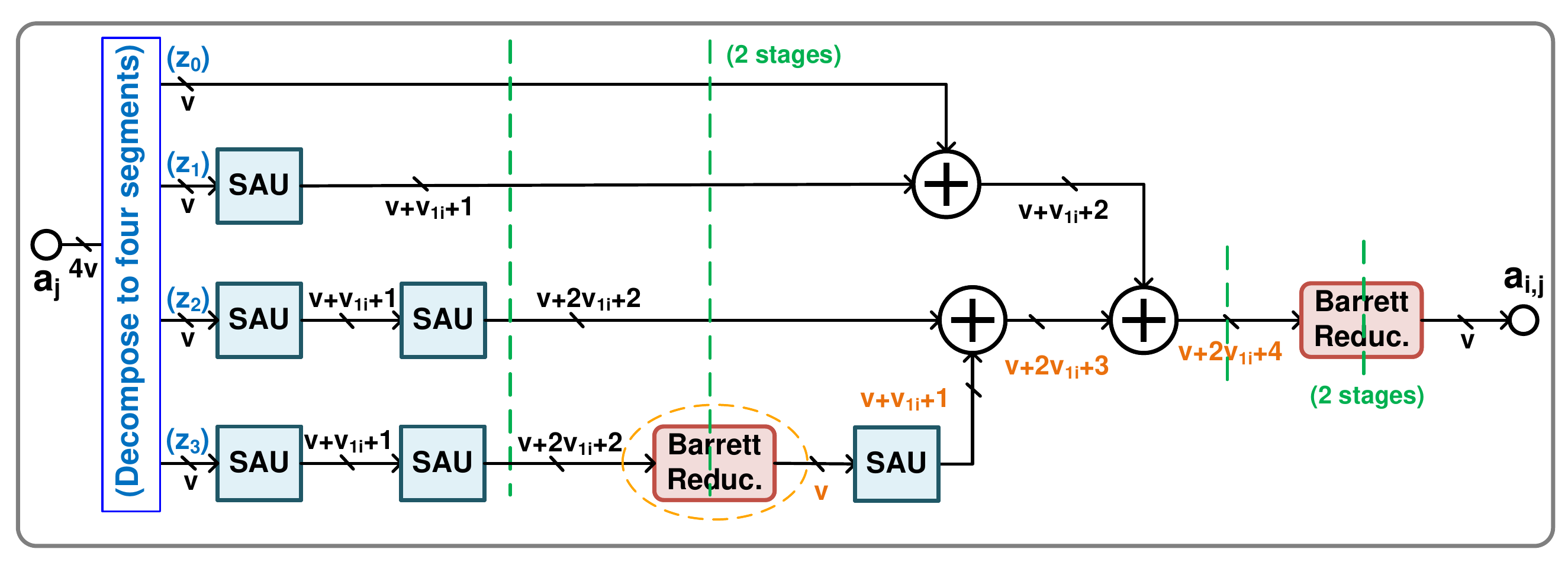}}
\caption{Residual coefficient computation unit with additional Barrett reduction unit for $t=4$. An additional Barrett reduction unit, employed to reduce the depth of the SAU, is circled, while word-lengths after the placement of this additional Barrett reduction unit are highlighted in orange.}
\figlbl{crt_pre_more_appro_1}

\end{figure}

In terms of computational complexity analysis, this proposed method demonstrates a reduction in hardware resource consumption. Compared to the design in \figref{archi-pre-pro}(a), this approach reduces the number of integer multipliers and modular reduction units from $t$ and $t$ to $(d-1)$ and $d$, respectively. However, additional $\frac{t(t'-1)}{2}$ SAUs are used. For example, the proposed method reduces six integer multipliers and six modular reduction units to one integer multiplier and two modular reduction units when $t=2\cdot 3$.
It is important to know that employing this method does not mandate the constraint parameter of $t=6$ for co-prime ($q_i$) generation during the exhaustive search procedure. On the contrary, it leverages the constraint parameter of $t'=3$ to achieve six satisfied co-prime factors since the maximum depth of SAU unit is two (i.e., $n_\beta = 2$ instead of $n_\beta = 5$), which markedly broadens the flexibility of the search space for co-prime factors.
\begin{figure}[htbp]
\centering
\resizebox{0.45\textwidth}{!}{
\includegraphics{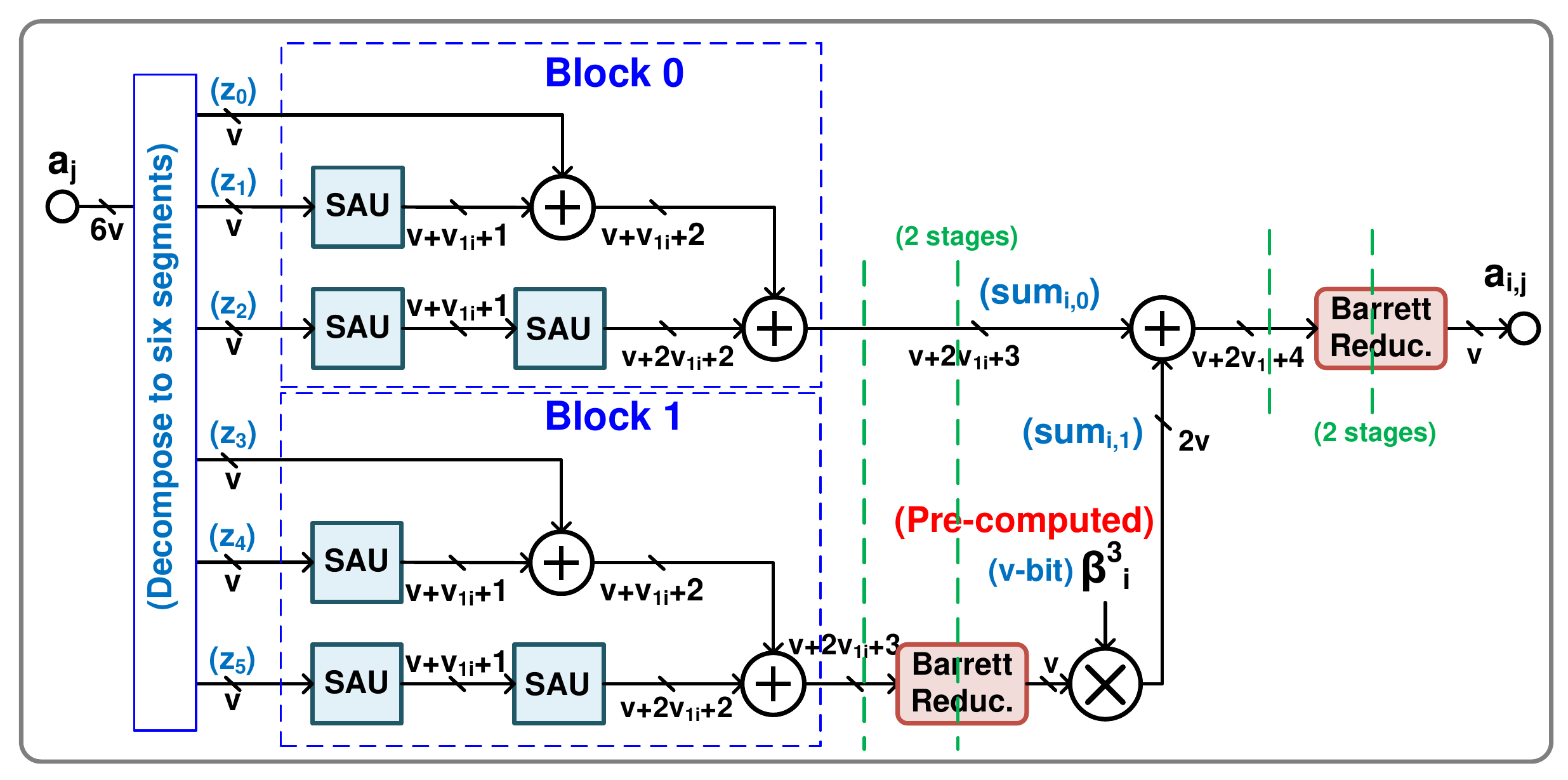}}
\caption{Residual coefficient computation by factorization unit for $t=6$. }
\figlbl{crt_pre_more}

\end{figure}
\subsection{Evaluation in residual domain}
After using CRT representation, the function $f(a_i(x),b_i(x))$ over $R_{n,q_i}$ can be computed independently, as presented in Step 2 in \figref{overview}. As a result, the overall $t$ operations can be executed in parallel. In our case, the function computes the residual products $p_i(x)$ for $i\in [1,t]$, by utilizing NTT-based polynomial multiplication over $R_{n,q_i}$. The architecture to compute $p_i(x) = a_i(x) \cdot b_i(x) \mod (x^n+1,q_i)$ is based on our novel NTT-based polynomial multiplier in \figref{polymult}. Thus, our proposed architecture achieves high throughput and low latency by increasing the parallelism from the CRT representation.  

\subsection{Inverse mapping of residual coefficients of polynomials}

During Step 3 in \figref{overview}, the results obtained by the evaluation in the residual domain need to be converted back to over the ring $R_{n,q}$, which is the same as $f(a(x),b(x))$ over $R_{n,q}$ (i.e., result computed without using CRT representation).

This post-processing stage is based on the inverse CRT algorithm:
\begin{align}
    p(x) &= \sum _{i=1}^t p_i(x) \cdot e_i \text{ mod } q \nonumber\\
    &= \sum _{i=1}^t \sum _{j=0}^{n-1} p_{i,j} \cdot e_i \cdot x^j \text{ mod } q,\eqnlbl{eq_icrt}
\end{align}
where each $e_i = q^{*}_{i} \cdot \tilde{q_i}$ is a constant, $q^{*}_{i} = (\frac{q}{q_i})\in \mathbb{Z}$, and $\tilde{q_i} = [(\frac{q}{q_i})^{-1}]_{q_i}\in \mathbb{Z}_{q_i}$.

However, direct multiplication by the constant $e_i$ involves a long integer multiplication and expensive modular reduction over $q$, which will result in an inefficient implementation and a long critical path. Meanwhile, the properties of the special co-primes can lower the cost of modular operations over $q_i$ in the post-processing stage. Therefore, we leverage the technique in~\cite{halevi2019improved} to further express \eqnref{eq_icrt} as: 
\begin{equation}
\begin{aligned}
    p(x) &= \sum _{i=1}^t \big[ p_i(x) \cdot \tilde{q_i}\big]_{q_i} \cdot q^{*}_{i} \text{ mod } q \\
    &= \sum _{i=1}^t \sum _{j=0}^{n-1} \big[p_{i,j} \cdot \tilde{q_i}\big]_{q_i} \cdot q^{*}_{i} \cdot x^j\text{ mod } q.
\end{aligned}
\end{equation}
The core concept of this methodology is the partitioning of a long word-length $v\times vt$-bit multiplier into a $v\times v$-bit multiplier coupled with a $v\times (t-1)v$-bit multiplier. Thus, the modular reduction with respect to $q$ is replaced by four separate modular reductions in terms of different $q_i$. The resource savings achieved through this optimization can be explained as follows:

The computation in $0 \leq \big[p_{i,j} \cdot \tilde{q_i}\big]_{q_i} < q_i$ can be performed efficiently since the modular reduction over $q_i$ has a lower cost than the modular reduction $q$. As $q^{*}_{i}$ is a $(t-1)v$-bit pre-computed constant, no division is required in the post-processing stage. Besides, the range of the coefficients from $\big[ p_i(x) \cdot \tilde{q_i}\big]_{q_i} \cdot q^{*}_{i}$ is in $[0,q-1]$ so that no modular multiplication is required to compute the product.  
\begin{figure}[htbp]
\centering
\resizebox{0.35\textwidth}{!}{
\includegraphics{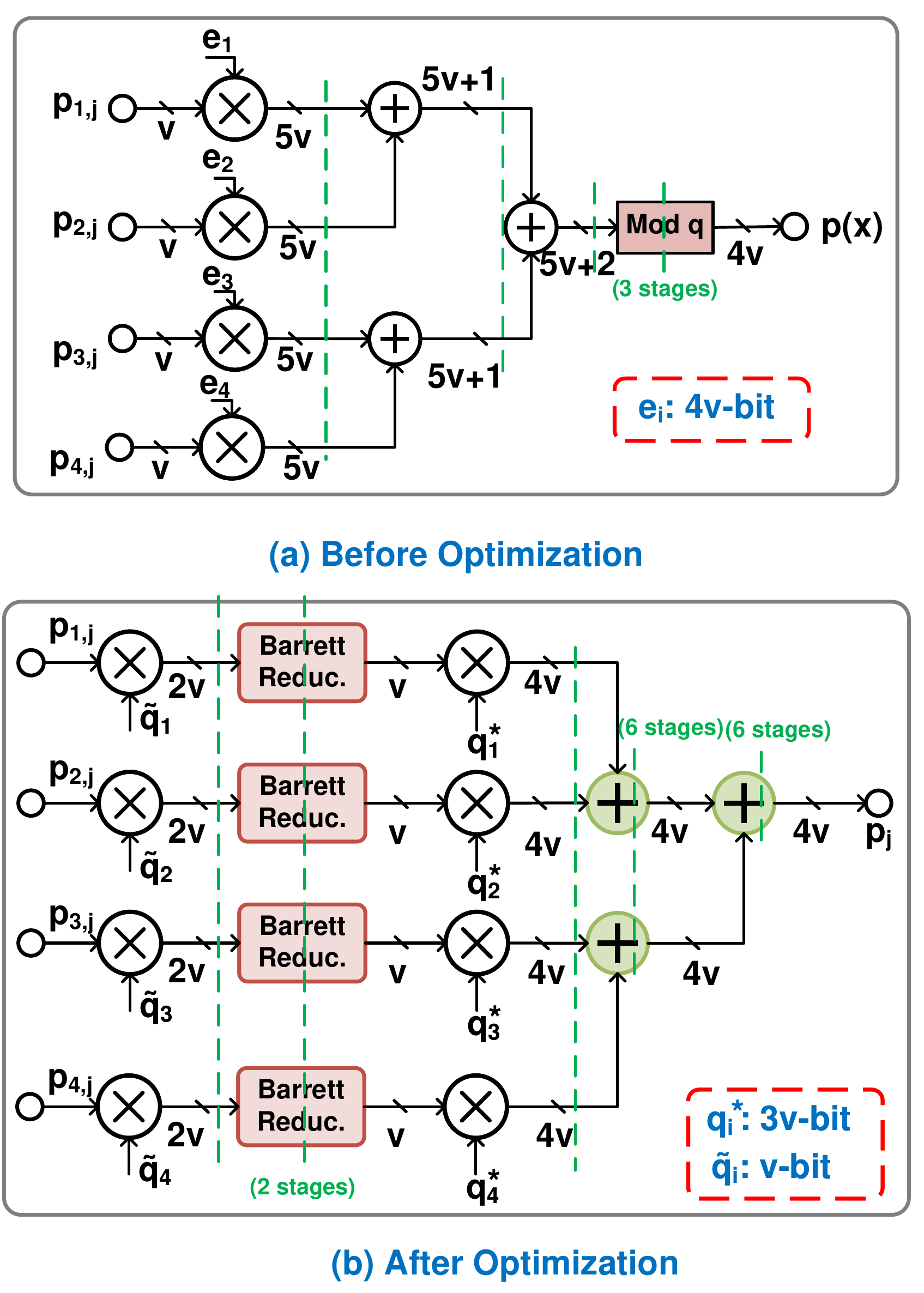}}
\caption{Inverse mapping architecture when $t = 4$. 
This circuit illustrates the post-processing step for the inverse CRT. Pipelining cut set is added after each integer adder/multiplier. }
\figlbl{archi_post_pro}

\end{figure}

The optimized architecture of the inverse mapping of residual coefficients of polynomials is shown in \figref{archi_post_pro}(b) (we use $t=4$ as an example). 
In this architecture, each long word-length ($4v\times v$-bit) multiplier for multiplying $e_i$ is split into $v\times v$-bit multiplier with constant $\tilde{q_i}$ and $v\times 2v$-bit multiplier with constant $q^*_i$. Instead of implementing an expensive modular reduction over a large modulus $q$ block in~\figref{archi_post_pro}(a), only three modular adders and four modular reductions over $q_i$ are required to obtain the final result $p(x)$. Specifically, the modular reduction over $q_i$ is also efficient based on the special co-prime. 

Overall, the proposed novel architecture can significantly reduce the area and power consumption.

\section{Experimental Results}\seclbl{result}
This section evaluates the co-prime factor selection and performance of the parallel NTT-based polynomial multiplier without shuffling operations (as presented \secref{sec_ntt}) and pre-processing/post-processing units for the CRT algorithm (detailed in \secref{sec_crt}) separately. Subsequently, a comprehensive performance discussion and comparison analysis of the proposed PaReNTT polynomial multiplier is presented.

For our evaluations, the proposed designs are implemented using SystemVerilog and then mapped to the Xilinx Virtex Ultrascale+ FPGA. We also specifically consider a fixed 180-bit $q$ with either four or six co-prime factors and a polynomial degree of $n=4096$ to investigate the designs under different levels of CRT-based parallelism. Consequently, the 180-bit modulus $q$ is composed of co-primes that are either 45-bit or 30-bit, and these co-primes adhere to special NTT-compatible and CRT-friendly formats.

Note that our design can be easily extended to a longer word-length modulus by either incorporating more co-prime factors or by increasing the word length of each individual co-prime.
Moreover, in the case of a length-$4096$ and $t=4$ NTT-based polynomial multiplier, 48 PEs and 44 DSD units are employed given that $m=\log_2 (4096)=12$. When $t=6$, 72 PEs and 66 DSD units are applied. 
A higher degree of the polynomial can also be integrated, requiring solely an increment in the number of PEs and DSDs.

\begin{figure}[htbp]
\centering
\resizebox{0.30\textwidth}{!}{
\includegraphics{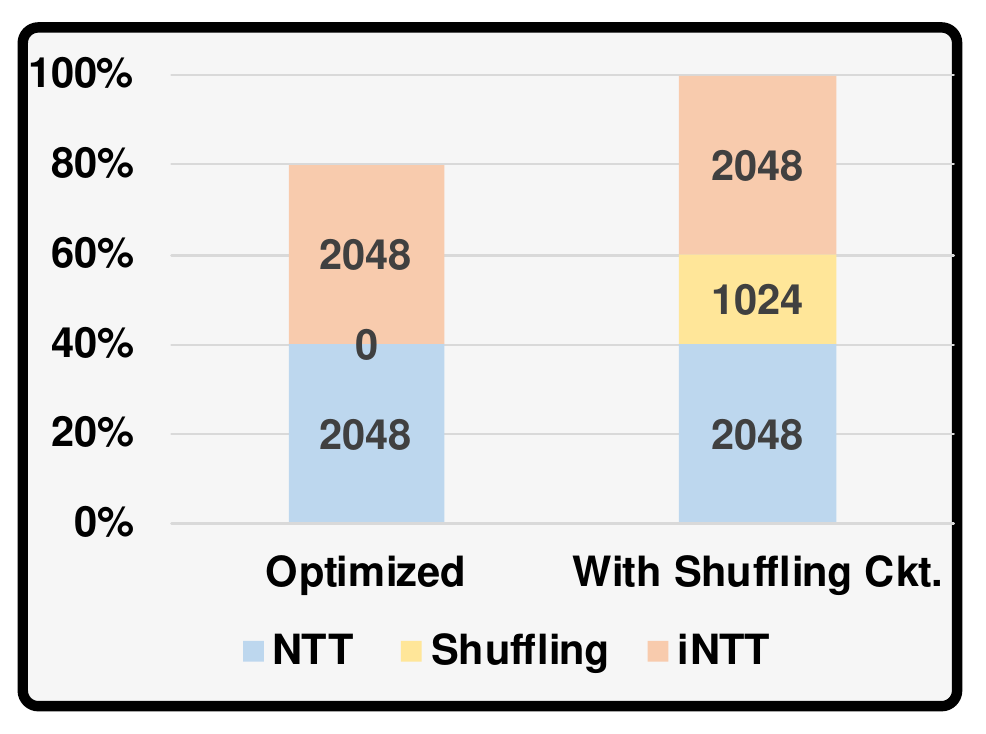}
}
\caption{Comparison of latency of Two-parallel NTT-based polynomial multiplication with and without shuffling operations when $n=4096$. }
\figlbl{ntt_bar}
\vspace{-1.5em}
\end{figure}
\subsection{Expansion of Co-prime factors}

Table \ref{tb_prime} shows the total number of special NTT-compatible and CRT-friendly primes under different settings. Two values of $\mu$ are chosen: $\mu ~=~ (2v+15)$ and $(2v~+~30)$. The number of signed power-of-two terms are either $4$ or $5$. When $\mu$, the number of signed power-of-two terms, and $n$ are set to be $(2v+30)$-bit, five terms, and length-4096, respectively, the feasible co-prime factors are 169 (for $v=30$) and 480 (for $v=45$) in number. Thus, the number of coprime factors is large enough to accommodate long word-lengths of coefficients. In our hardware implementation, 75-bit and 105-bit $\mu$ are considered for the 30-bit ($v=30$) and 45-bit ($v=45$) co-primes (corresponding to $\mu = (2v+15)$). Moreover, each of these co-primes is characterized by four signed power-of-two terms.
\begin{table}[htbp]
\centering
\begin{threeparttable}[htbp]
  \caption{The number of special NTT-compatible and CRT-friendly primes under different settings when $t=4$ and $t=6$ ($v = 45$ and 30)}\label{tb_prime}
  \setlength{\tabcolsep}{6pt}
{\begin{tabular}{|c|c|c|c|c|c||c|}
\hline
 $t$ & $v$ &$\mu$ & \small{\# PoT}  &  $n$  & \small{$\lceil \log_2 \epsilon \rceil $} &\small{\# primes}\\
 \hline
4 & 45 &$(2v+15)$ & 4 & 4096 & 61 &12 \\
 \hline
4 & 45 &$(2v+30)$ & 4 & 4096 & 76 &33 \\
\hline
4 & 45 &$(2v+15)$ & 5 & 4096 & 61 &126 \\
 \hline
4 & 45 &$(2v+30)$ & 5 & 4096 & 76 &480 \\
\hline
6 & 30 &$(2v+15)$ & 4 & 4096 & 46 &8 \\
 \hline
6 & 30 &$(2v+30)$ & 4 & 4096 & 61 &26 \\
\hline
6 & 30 &$(2v+15)$ & 5 & 4096 & 46 &23 \\
 \hline
6 & 30 &$(2v+30)$ & 5 & 4096 & 61 &169 \\
\hline
\end{tabular}
}  
\begin{tablenotes}
     \small
     \item  {$\mu$: The input word-length of Barrett reduction unit; \# PoT: The number of signed power-of-two terms in each co-prime. }
\end{tablenotes}
\end{threeparttable}
\end{table}

\subsection{Evaluation metrics and performance of parallel NTT-based polynomial multiplier}
To analyze the timing performances of the implementations, we define two timing performance metrics, {\em block processing period} (BPP) and latency. BPP is defined as the time required to process $n$ coefficient inputs or the time required to generate $n$ coefficient outputs. For a length-$n$ NTT-based two-parallel polynomial multiplier, the expression for BPP is 
\begin{equation}
    T_{BPP} = n/2,
\end{equation}
where the throughput is two samples per clock cycle.
In addition, the latency for one modular polynomial multiplication is
\begin{equation}
    T_{Lat} = (n - 2) + T_{pipe},
\end{equation}
where $T_{pipe}$ represents the additional pipelining stages added to the data-path in order to reduce the critical path. Furthermore, the total clock cycles consumed by $L$ modular polynomial multiplications are
\begin{equation}
    T_{total} = T_{Lat} + T_{BPP} \cdot L.
\end{equation}

For $n=4096$, the BPP is 2048 clock cycles, and the latency is 4096 clock cycles (excluding extra clock cycles required for pipelining). The latency is significantly reduced compared to the NTT-based polynomial multipliers that use a shuffling circuit in the prior works. The comparison of our optimized and conventional methods (without considering the pipelining) is shown in \figref{ntt_bar}. Specifically, the conventional method with the shuffling circuit needs additional 1024 ($n/4$ in general) clock cycles for the re-ordering, leading to an increase in latency by around $20.0\%$ for a two-parallel design and $n~=~4096$. 
\begin{table}[htbp]
\centering
  \caption{Area consumption and frequency for residual coefficient computation unit when $t=4$ and $t=6$ ($\lceil \log_2 q_i\rceil = 45$ and 30)}\label{tb_crt_area}
  \resizebox{0.4\textwidth}{!}
{\begin{tabular}{|c||c|c|c|c|c|c|}
\hline
Design &$t$ & Freq. & LUTs  &  DSPs  & FFs &$N_{pip}$\\
 \hline
Prior work &4 & 76 & 6350 & 0 & 0 &0 \\
 \hline
\textbf{Proposed}  &4 & \textbf{62} & \textbf{4034} &\textbf{0}&\textbf{0}&\textbf{0}\\
 \hline

Prior work &4 & 200 & 5836 & 0 & 1288 &5 \\
 \hline
\textbf{Proposed}  &4 &\textbf{271}  &\textbf{3937} &\textbf{0} & \textbf{1164}  &\textbf{6}\\
 \hline
Prior work &6 & 105 & 2032 & 0 & 0 &0 \\
 \hline
\textbf{Proposed}  &6 & \textbf{55} & \textbf{1148} &\textbf{0}&\textbf{0}&\textbf{0}\\
 \hline

Prior work  &6 & 300 & 2660 & 0 & 1244 &6 \\
 \hline
\textbf{Proposed}  &6 &\textbf{309}  &\textbf{1537} &\textbf{0} & \textbf{682}  &\textbf{6}\\
 \hline
\end{tabular}
}  
\vspace{-1.5em}
\end{table}

\begin{table}[htbp]
\centering
  \caption{Area consumption and frequency for  inverse mapping architecture when $t=4$ and $\lceil \log_2 q_i\rceil = 45$}\label{tb_icrt_area}
  \resizebox{0.4\textwidth}{!}
{\begin{tabular}{|c||c|c|c|c|c|}
\hline
Design  & Freq. & LUTs  &  DSPs  & FFs &$N_{pip}$\\
 \hline
Conven.&45  &17729   & 63  &0 &0 \\
 \hline
\textbf{Proposed} & \textbf{50} & \textbf{15894} &\textbf{60}& \textbf{0} &\textbf{0}\\
 \hline
Conven.&111  &15066   & 63  &2544 &6 \\
 \hline
\textbf{Proposed} & \textbf{244} & \textbf{12302} &\textbf{60} & \textbf{6686} &\textbf{16}\\
 \hline
\end{tabular}
}  
\vspace{-2em}
\end{table}

\begin{table*}[htbp]
\centering
\begin{threeparttable}[htbp]
  \caption{Area consumption and frequency for modular polynomial multipliers for $n=4096$}\label{tb_FPGA_area_4096}
  \setlength{\tabcolsep}{10pt}
{\begin{tabular}{|c|c|c|c|c|c|c|c|c|c|c|c|}
\hline
Design &$\lceil \log_2 q\rceil$ & $t$   & Freq.[MHz] & LUTs$^a$ &  DSPs$^a$ & FFs$^a$ & Power [W]\\

 \hline
\multirow{2}{*}{Proposed} & 180 & 4  & 244 & 322K (27.2\%) & 1.6K (22.8\%) & 92K (3.9\%) &6.6\\
 \cline{2-8}
                          & 180 & 6  & 240 & 341K (28.9\%) & 1.1K (16.5\%) & 103K (4.3\%) &6.3\\
 \hline
Roy~\cite{roy2019fpga}& 180 &6  & 225 & 64K & 0.3K & 25K & \small{(Not Reported)}\\
 \hline
\end{tabular}
}
\begin{tablenotes}
     \small
     \item  {$^a$: \# of used resources (\% utilization) on FPGA board.}
\end{tablenotes}
\end{threeparttable}
\vspace{-1.5em}
\end{table*}

\begin{table*}[htbp]
\begin{center}
\begin{threeparttable}[htbp]
  \caption{Timing performance  for modular polynomial multipliers for $n=4096$}\label{tb_FPGA_timing_4096}
  
  \setlength{\tabcolsep}{6pt}
{\begin{tabular}{|c|c|c|c|c|c|c|c|c|c|c|c|c|}
\hline
\multirow{2}{*}{$\lceil \log_2 q\rceil$} & \multirow{2}{*}{ $t$ } & \multirow{2}{*}{CRT} & \multicolumn{2}{c|}{\multirow{1}{*}{BPP}$^b$} & \multicolumn{2}{c|}{\multirow{1}{*}{Latency}$^c$} &{ABP$^d$} &{ABP$^d$} &{ATP$^e$} &{ATP$^e$} \\
\cline{4-7}
 & & & \multicolumn{1}{c|}{$\#$ Cycles} & \multicolumn{1}{c|}{Period [$\mu$s]} & \multicolumn{1}{c|}{$\#$ Cycles} & \multicolumn{1}{c|}{Period [$\mu$s]} & (LUT) & (DSP) & (LUT) & (DSP)\\
 \hline
 180 & 4 & Yes & 2048 & 8.5 & 4246 & 17.4 & 2.7M & 13.1K & 5.6M & 27.8K \\
 \hline
 180 & 6 & Yes & 2048 & 8.4 & 4254 & 17.7 & 2.9M & 9.6K & 6.0M & 19.5K\\
 \hline
  180 & 6 & Yes & N/A & N/A & 196003 & 871.1 & N/A & N/A & 55.8M & 261.3K\\
 \hline
\end{tabular}
}
\begin{tablenotes}
     \small
     \item  {$^b$: Block processing period (BPP) is the period ($\mu$s) for processing $n$ coefficient inputs or for generating $n$ sample outputs after the first sample out.}
     \item  {$^c$: Latency is the period ($\mu$s) of the first sample in and the first sample out.}
     \item  {$^d$: ABP is calculated from the number of LUTs/DSPs times BPP ($\mu$s).} 
     \item  {$^e$: ATP is calculated from the number of LUTs/DSPs times latency ($\mu$s).} 
\end{tablenotes}
\end{threeparttable}
\end{center}
\vspace{-2em}
\end{table*}
\subsection{Comparison of residual coefficient computation unit and inverse mapping architecture}
\figref{crt_pre_more_appro_1} and \figref{crt_pre_more} illustrate and compare the designs for residual coefficient, and \figref{archi_post_pro} presents inverse mapping computations architecture.  The experimental results and comparison, both with and without the incorporation of pipelining for these foundational components, are presented in Tables \ref{tb_crt_area} and \ref{tb_icrt_area}. The pipelining cut sets in the building blocks are marked in green in \figref{crt_pre_more_appro_1}, \figref{crt_pre_more}, and \figref{archi_post_pro}.

In evaluating the results for our proposed residual coefficient computation unit, we have considered the experimental results for two distinct approaches presented in \figref{crt_pre_more_appro_1} (for $v=45$, $t=4$) and \figref{crt_pre_more} (for $v=30$, $t=6$). Additionally, we reference the prior design delineated in \figref{archi-pre-pro}(a) and implemented it in a fully parallel manner with our parameter setting and Barrett reduction units. This has been employed as a baseline for the comparison of the residual coefficient computation unit presented in Table~\ref{tb_crt_area}. Both pipelined and non-pipelined designs are considered. From non-pipelined designs ($N_{pip}~=~0$), we observe that the area requirements of the proposed designs for preprocessing are less than those of the prior design. A comparison between the prior design (\figref{archi-pre-pro}(a)) and the proposed design of \figref{crt_pre_more_appro_1} reveals a shorter critical path in the former before pipelining. These designs are feed-forward and can be pipelined at appropriate levels. For a fair comparison, both designs are appropriately pipelined to facilitate high-speed operation.
The result indicates a significant reduction of 32.5\% in LUT consumption in our design. Such saving mainly comes from replacing the four integer multipliers and four Barrett reduction units by two Barrett reduction units augmented by additional low-cost SAU units. 
Meanwhile, the comparison between \figref{archi-pre-pro}(a) and our proposed design in \figref{crt_pre_more} shows a saving of LUTs increases to $67.7\%$ after pipelining. 

Besides, parameter setting of $v=45$ and $t=4$ is applied to compare conventional design \figref{archi_post_pro}(a) and our proposed design \figref{archi_post_pro}(b) for the inverse mapping architecture. The area consumption results show an 18.3\% and 4.8\% reduction in the usage of LUTs and DSPs in our proposed design, respectively.  Such savings are primarily derived from the replacement of an expensive Barrett reduction unit with respect to $q$ by four Barrett units using special primes $q_i$. In particular, instead of performing a multiplication with a 180-bit integer $q$ during the Barrett reduction with $q$, our approach employs four short word-length shift-and-add operations to compute the multiplications with 45-bit specialized $q_i$. Although the total word-length of multipliers for $\tilde{q_i}$ and $q^*_i$ remains unchanged compared to the multiplier with $e_i$, the decomposition of the long word-length multiplication at the algorithmic level for our proposed hardware architecture enables a straightforward pipelining optimization without the need for further transformation.

\subsection{Evaluation on PaReNTT polynomial multiplier}
This sub-section delves into the implementation and comparison of the  proposed PaReNTT polynomial multiplier (two-parallel residue arithmetic-based NTT architecture) for $n=4096$ and $\lceil \log_2 q\rceil = 180$. 

The performances and experimental results for the parameter settings $t=4$, $v=45$ and $t=6$, $v=30$ are presented in Tables  \ref{tb_FPGA_area_4096} and \ref{tb_FPGA_timing_4096}. 
These two implementations employ the same architecture designs for the evaluation in the residual domain (i.e., the parallel NTT-based polynomial multiplier for varying $q_i$ as described in \secref{sec_ntt}) and the inverse mapping of residual coefficients of the polynomial. 
However, the employed residual polynomial computation units for $t=4$ and $t=6$ are based on \figref{crt_pre_more_appro_1} and \figref{crt_pre_more}, respectively. Detailed breakdowns of these two blocks' results are presented in Table \ref{tb_crt_area}.

In terms of timing performance, both designs can operate at a high clock frequency of 240MHz after pipelining. It can also be observed that the BPP and latency, measured in clock cycles, remain similar regardless of the varying word-length $v$ due to the degree of the polynomial being fixed. Furthermore, the area performance of PaReNTT architectures for $t=4$ and $t=6$ is also examined. As illustrated in Table \ref{tb_FPGA_area_4096}, the implementation for $t=6$ utilizes an additional 5.6\% of LUTs, while concurrently reducing DSP usage by 31.25\% compared to the design implemented for $t=4$. 

To comprehensively compare the timing and area performances of our proposed designs, we evaluate the area-BPP product (ABP). The reductions in ABP(LUT) and ABP(DSP) achieved by the $t=6$ design are 6.90\% and 26.72\%, respectively, when compared to the $t=4$ design.

The main sources of power consumption in our PaReNTT architectures are the shift registers deployed in the DSD units, in addition to the logic operations executed in LUTs and DSPs. Since the $t=6$ implementation utilizes fewer resources, it is associated with a reduction in power consumption. Specifically, it is approximately 4.5\% lower when compared to the $t=4$ implementation.

Although the parameter setting of $n=4096$ and $\lceil \log_2 q\rceil = 180$ indicates superior ABP(LUT) and ABP(DSP) performance for the $t=6$ implementation, varying parameter selections for $n$, $v$, and $t$ may also impact both the flexibility of co-prime factor selection and ABP performance. This suggests that the choice between designs shown in \figref{crt_pre_more_appro_1} or \figref{crt_pre_more} and the selection of parameters should be meticulously tailored to suit the requirements of different HE applications.

Direct comparisons with prior works are difficult as systems are implemented using different data-paths and FPGA devices corresponding to different technologies. Nevertheless, we now compare the proposed design with a prior design based on the same parameter setting $n=4096$, $\log_2(q) = 180$ in~\cite{roy2019fpga} and the same FPGA device. The timing and area performances of the prior design are included in the last line of Tables \ref{tb_FPGA_timing_4096} and \ref{tb_FPGA_area_4096}. Moreover, to reduce the variation of the parameter setting, parameter setting of $v=30$ and $t=6$ is considered in the proposed PaReNTT architecture, which is the same as \cite{roy2019fpga}.

Despite the fact that the area performance of the previous design is superior to the PaReNTT architecture, our design has a better timing performance, as reducing the latency and increasing throughput is the primary goal of this work. Specifically, the prior design incorporates a customized optimization for the BFV scheme requiring lifting and scaling operations. 
Consequently, the clock cycles for modular multiplication in the homomorphic multiplication are approximately doubled compared to the design without these operations. In order to provide a fair comparison, we halved the clock cycle and latency consumption for the CRT-based, NTT, and iNTT operations in their design. The equivalent number of clock cycles equals $196,003 = (87,582\times 2+102,043+15,662+99,137)/2$, and the latency is  $871.1$ $\mu$s. Note that the approximated timing results are obtained from Table II in~\cite{roy2019fpga} by calculating the sum of two NTTs (for polynomial $a(x)$ and $b(x)$, respectively), coefficient-wise multiplication, iNTT, Lift and then divided by two. As their scaling step requires a more complex operation than the general inverse mapping for the residual polynomials due to scheme requirements, the clock cycles in this step are excluded from their approximated timing result.

The comparison and evaluation result shows our design reduces the latency by a factor of 49.2. Additionally, we compare the area-timing product (ATP) of these two designs. The comparison indicates that our design reduces ATP(LUT) and ATP(DSP) by 89.2\% and 92.5\%, respectively, compared to the design in \cite{roy2019fpga}.
\vspace{-1.5em}

\section{Conclusion}\seclbl{conclusion}
This paper has proposed PaReNTT, an efficient CRT and NTT-based long polynomial multiplier. This design leverages the characteristics of the specially selected primes to optimize the pre-processing and post-processing units for the CRT algorithm. In addition, a novel iNTT unit is designed based on bit-reversed scheduling to eliminate an expensive shuffling circuit and significantly reduce latency. Future work will be directed toward evaluating different homomorphic encryption algorithms such as BFV, BGV, and CKKS using the proposed efficient long polynomial multiplier based on hardware-software co-design. 
\vspace{-1.5em}

\bibliographystyle{IEEEtran}
% \balance
\bibliography{main}

\begin{thebibliography}{10}
\providecommand{\url}[1]{#1}
\csname url@samestyle\endcsname
\providecommand{\newblock}{\relax}
\providecommand{\bibinfo}[2]{#2}
\providecommand{\BIBentrySTDinterwordspacing}{\spaceskip=0pt\relax}
\providecommand{\BIBentryALTinterwordstretchfactor}{4}
\providecommand{\BIBentryALTinterwordspacing}{\spaceskip=\fontdimen2\font plus
\BIBentryALTinterwordstretchfactor\fontdimen3\font minus
  \fontdimen4\font\relax}
\providecommand{\BIBforeignlanguage}[2]{{%
\expandafter\ifx\csname l@#1\endcsname\relax
\typeout{** WARNING: IEEEtran.bst: No hyphenation pattern has been}%
\typeout{** loaded for the language `#1'. Using the pattern for}%
\typeout{** the default language instead.}%
\else
\language=\csname l@#1\endcsname
\fi
#2}}
\providecommand{\BIBdecl}{\relax}
\BIBdecl

\bibitem{wibawa2022homomorphic}
F.~Wibawa, F.~O. Catak, M.~Kuzlu, S.~Sarp, and U.~Cali, ``Homomorphic
  encryption and federated learning based privacy-preserving {CNN} training:
  {COVID-19} detection use-case,'' in \emph{Proceedings of the 2022 European
  Interdisciplinary Cybersecurity Conference}, 2022, pp. 85--90.

\bibitem{chen2019efficient}
H.~Chen, W.~Dai, M.~Kim, and Y.~Song, ``Efficient multi-key homomorphic
  encryption with packed ciphertexts with application to oblivious neural
  network inference,'' in \emph{Proceedings of the 2019 ACM SIGSAC Conference
  on Computer and Communications Security}, 2019, pp. 395--412.

\bibitem{lyubashevsky2010ideal}
V.~Lyubashevsky, C.~Peikert, and O.~Regev, ``On ideal lattices and learning
  with errors over rings,'' in \emph{Annual International Conference on the
  Theory and Applications of Cryptographic Techniques}.\hskip 1em plus 0.5em
  minus 0.4em\relax Springer, 2010, pp. 1--23.

\bibitem{halevi2014algorithms}
S.~Halevi and V.~Shoup, ``Algorithms in {helib},'' in \emph{Annual Cryptology
  Conference}.\hskip 1em plus 0.5em minus 0.4em\relax Springer, 2014, pp.
  554--571.

\bibitem{chen2017simple}
H.~Chen, K.~Laine, and R.~Player, ``Simple encrypted arithmetic library-{SEAL}
  v2.1,'' in \emph{International Conference on Financial Cryptography and Data
  Security}.\hskip 1em plus 0.5em minus 0.4em\relax Springer, 2017, pp. 3--18.

\bibitem{roy2018hepcloud}
S.~S. Roy, K.~Jarvinen, J.~Vliegen, F.~Vercauteren, and I.~Verbauwhede,
  ``{HEPCloud}: An {FPGA}-based multicore processor for {FV} somewhat
  homomorphic function evaluation,'' \emph{IEEE Transactions on Computers},
  2018.

\bibitem{roy2019fpga}
S.~S. Roy, F.~Turan, K.~Jarvinen, F.~Vercauteren, and I.~Verbauwhede,
  ``{FPGA}-based high-performance parallel architecture for homomorphic
  computing on encrypted data,'' in \emph{2019 IEEE International Symposium on
  High Performance Computer Architecture (HPCA)}.\hskip 1em plus 0.5em minus
  0.4em\relax IEEE, 2019, pp. 387--398.

\bibitem{riazi2020heax}
M.~S. Riazi, K.~Laine, B.~Pelton, and W.~Dai, ``{HEAX}: An architecture for
  computing on encrypted data,'' in \emph{Proceedings of the Twenty-Fifth
  International Conference on Architectural Support for Programming Languages
  and Operating Systems}, 2020, pp. 1295--1309.

\bibitem{gentry2009fully}
C.~Gentry, \emph{A fully homomorphic encryption scheme}.\hskip 1em plus 0.5em
  minus 0.4em\relax Stanford university, 2009.

\bibitem{fan2012somewhat}
J.~Fan and F.~Vercauteren, ``Somewhat practical fully homomorphic encryption.''
  \emph{IACR Cryptology ePrint Archive}, vol. 2012, p. 144, 2012.

\bibitem{cheon2017homomorphic}
J.~H. Cheon, A.~Kim, M.~Kim, and Y.~Song, ``Homomorphic encryption for
  arithmetic of approximate numbers,'' in \emph{International Conference on the
  Theory and Application of Cryptology and Information Security}.\hskip 1em
  plus 0.5em minus 0.4em\relax Springer, 2017, pp. 409--437.

\bibitem{aysu2013low}
A.~Aysu, C.~Patterson, and P.~Schaumont, ``Low-cost and area-efficient {FPGA}
  implementations of lattice-based cryptography,'' in \emph{2013 IEEE
  International Symposium on Hardware-Oriented Security and Trust
  (HOST)}.\hskip 1em plus 0.5em minus 0.4em\relax IEEE, 2013, pp. 81--86.

\bibitem{dai2015cuhe}
W.~Dai and B.~Sunar, ``{cuHE}: A homomorphic encryption accelerator library,''
  in \emph{International Conference on Cryptography and Information Security in
  the Balkans}.\hskip 1em plus 0.5em minus 0.4em\relax Springer, 2015, pp.
  169--186.

\bibitem{lyubashevsky2008swifft}
V.~Lyubashevsky, D.~Micciancio, C.~Peikert, and A.~Rosen, ``{SWIFFT}: A modest
  proposal for {FFT} hashing,'' in \emph{International Workshop on Fast
  Software Encryption}.\hskip 1em plus 0.5em minus 0.4em\relax Springer, 2008,
  pp. 54--72.

\bibitem{zhang2020highly}
N.~Zhang, B.~Yang, C.~Chen, S.~Yin, S.~Wei, and L.~Liu, ``Highly efficient
  architecture of {NewHope-NIST} on {FPGA} using low-complexity {NTT/INTT},''
  \emph{IACR Transactions on Cryptographic Hardware and Embedded Systems}, pp.
  49--72, 2020.

\bibitem{longa2016speeding}
P.~Longa and M.~Naehrig, ``Speeding up the number theoretic transform for
  faster ideal lattice-based cryptography,'' in \emph{International Conference
  on Cryptology and Network Security}.\hskip 1em plus 0.5em minus 0.4em\relax
  Springer, 2016, pp. 124--139.

\bibitem{chiu2023nttbased}
S.-W. Chiu and K.~K. Parhi, ``{NTT}-based polynomial modular multiplication for
  homomorphic encryption: A tutorial,'' \emph{arXiv:2306.12519}, 2023.

\bibitem{xin2021multi}
G.~Xin, Y.~Zhao, and J.~Han, ``A multi-layer parallel hardware architecture for
  homomorphic computation in machine learning,'' in \emph{2021 IEEE
  International Symposium on Circuits and Systems (ISCAS)}.\hskip 1em plus
  0.5em minus 0.4em\relax IEEE, 2021, pp. 1--5.

\bibitem{turan2020heaws}
F.~Turan, S.~S. Roy, and I.~Verbauwhede, ``{HEAWS}: An accelerator for
  homomorphic encryption on the amazon aws fpga,'' \emph{IEEE Transactions on
  Computers}, vol.~69, no.~8, pp. 1185--1196, 2020.

\bibitem{xing2021compact}
Y.~Xing and S.~Li, ``A compact hardware implementation of {CCA}-secure key
  exchange mechanism {CRYSTALS-KYBER} on {FPGA},'' \emph{IACR Transactions on
  Cryptographic Hardware and Embedded Systems}, pp. 328--356, 2021.

\bibitem{mert2022medha}
A.~C. Mert, S.~Kwon, Y.~Shin, D.~Yoo, Y.~Lee, S.~S. Roy \emph{et~al.}, ``Medha:
  Microcoded hardware accelerator for computing on encrypted data,''
  \emph{Cryptology ePrint Archive}, 2022.

\bibitem{mert2020fpga}
A.~C. Mert, E.~{\"O}zt{\"u}rk, and E.~Sava{\c{s}}, ``{FPGA} implementation of a
  run-time configurable {NTT}-based polynomial multiplication hardware,''
  \emph{Microprocessors and Microsystems}, vol.~78, p. 103219, 2020.

\bibitem{paludo2022ntt}
R.~Paludo and L.~Sousa, ``{NTT} architecture for a {Linux}-ready {RISC-V}
  fully-homomorphic encryption accelerator,'' \emph{IEEE Transactions on
  Circuits and Systems I: Regular Papers}, 2022.

\bibitem{ayinala2011pipelined}
M.~Ayinala, M.~Brown, and K.~K. Parhi, ``Pipelined parallel {FFT} architectures
  via folding transformation,'' \emph{IEEE Transactions on Very Large Scale
  Integration Systems}, vol.~20, no.~6, pp. 1068--1081, 2011.

\bibitem{ayinala2013place}
M.~Ayinala, Y.~Lao, and K.~K. Parhi, ``An in-place {FFT} architecture for
  real-valued signals,'' \emph{IEEE Transactions on Circuits and Systems II:
  Express Briefs}, vol.~60, no.~10, pp. 652--656, 2013.

\bibitem{parhi2007vlsi}
K.~K. Parhi, \emph{{VLSI} digital signal processing systems: design and
  implementation}.\hskip 1em plus 0.5em minus 0.4em\relax John Wiley \& Sons,
  1999.

\bibitem{parhi1992synthesis}
K.~K. Parhi, C.-Y. Wang, and A.~P. Brown, ``Synthesis of control circuits in
  folded pipelined {DSP} architectures,'' \emph{IEEE Journal of Solid-State
  Circuits}, vol.~27, no.~1, pp. 29--43, 1992.

\bibitem{tan2021pipelined}
W.~Tan, A.~Wang, Y.~Lao, X.~Zhang, and K.~K. Parhi, ``Pipelined high-throughput
  {NTT} architecture for lattice-based cryptography,'' in \emph{2021 Asian
  Hardware Oriented Security and Trust Symposium (AsianHOST)}.\hskip 1em plus
  0.5em minus 0.4em\relax IEEE, 2021, pp. 1--4.

\bibitem{zhang2020efficient}
Y.~Zhang, C.~Wang, D.~E.~S. Kundi, A.~Khalid, M.~O’Neill, and W.~Liu, ``An
  efficient and parallel {R-LWE} cryptoprocessor,'' \emph{IEEE Transactions on
  Circuits and Systems II: Express Briefs}, vol.~67, no.~5, pp. 886--890, 2020.

\bibitem{tan2021high}
W.~Tan, B.~M. Case, A.~Wang, S.~Gao, and Y.~Lao, ``High-speed modular
  multiplier for lattice-based cryptosystems,'' \emph{IEEE Transactions on
  Circuits and Systems II: Express Briefs}, vol.~68, no.~8, pp. 2927--2931,
  2021.

\bibitem{hamburg2015ed448}
M.~Hamburg, ``Ed448-goldilocks, a new elliptic curve.'' \emph{IACR Cryptol.
  ePrint Arch.}, vol. 2015, p. 625, 2015.

\bibitem{barrett1986implementing}
P.~Barrett, ``Implementing the rivest shamir and adleman public key encryption
  algorithm on a standard digital signal processor,'' in \emph{Conference on
  the Theory and Application of Cryptographic Techniques}.\hskip 1em plus 0.5em
  minus 0.4em\relax Springer, 1986, pp. 311--323.

\bibitem{halevi2019improved}
S.~Halevi, Y.~Polyakov, and V.~Shoup, ``An improved {RNS} variant of the {BFV}
  homomorphic encryption scheme,'' in \emph{Cryptographers’ Track at the RSA
  Conference}.\hskip 1em plus 0.5em minus 0.4em\relax Springer, 2019, pp.
  83--105.

\end{thebibliography}
\begin{IEEEbiography}[{\includegraphics[width=1in,height=1.25in,clip]{TAN.pdf}}]{Weihang Tan} (Graduate Student Member, IEEE) 
is a postdoctoral research associate in the Department of Electrical and Computer Engineering at the University of Minnesota, Twin Cities. He received his B.S., M.S., and Ph.D. degrees in Electrical Engineering from Clemson University, Clemson, SC, USA, in 2018, 2020, and 2022, respectively. His research interests include hardware security and VLSI architecture design for fully homomorphic encryption, post-quantum cryptography, and digital signal processing systems. He is the recipient of the best PhD forum presentation award at the Asian Hardware Oriented Security and Trust Symposium (AsianHOST).
\end{IEEEbiography} 

\begin{IEEEbiography}[{\includegraphics[width=1in,height=1.33in,clip]{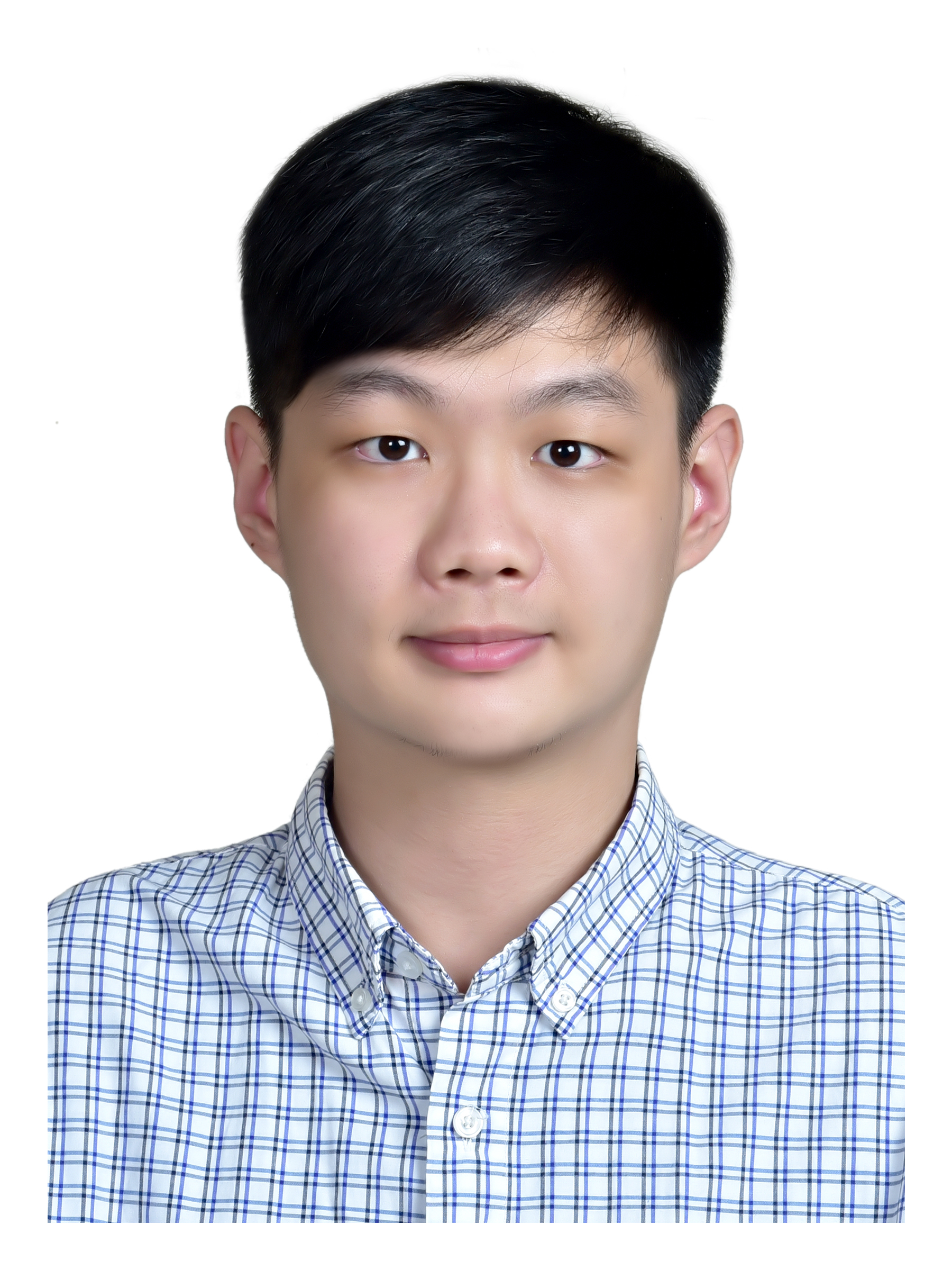}}]{Sin-Wei Chiu} (Graduate Student Member, IEEE) 
received his bachelor’s degree in electrical engineering from National Central University, Taiwan, in 2020. He is currently pursuing a Ph.D. degree in electrical engineering at the University of Minnesota, Twin Cities. His current research interests include VLSI architecture design, post-quantum cryptography, and digital signal processing systems. 
\end{IEEEbiography}

\begin{IEEEbiography}[{\includegraphics[width=1in,height=1.25in,clip,keepaspectratio]{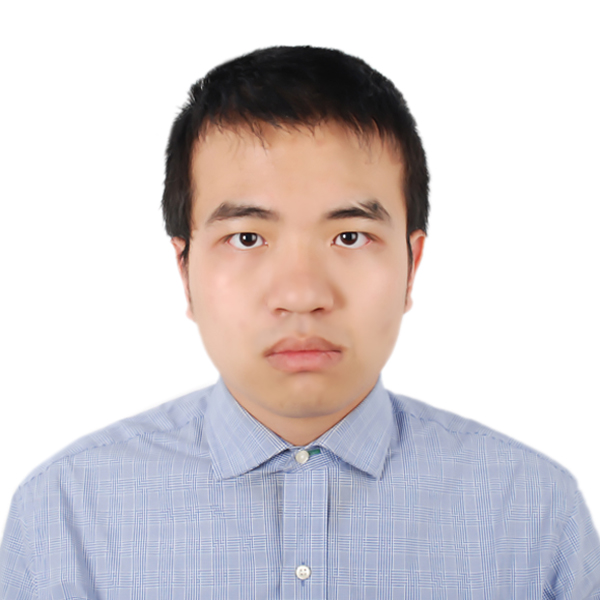}}]{Antian Wang} (Graduate Student Member, IEEE) 
received his B.E. (2017) in Communication Engineering from Shanghai Maritime University. He is currently pursuing a Ph.D. degree in Clemson University. His research interests include hardware security and VLSI architecture design, and design automation.
\end{IEEEbiography} 

\begin{IEEEbiography}[{\includegraphics[width=1in,height=1.25in,clip]{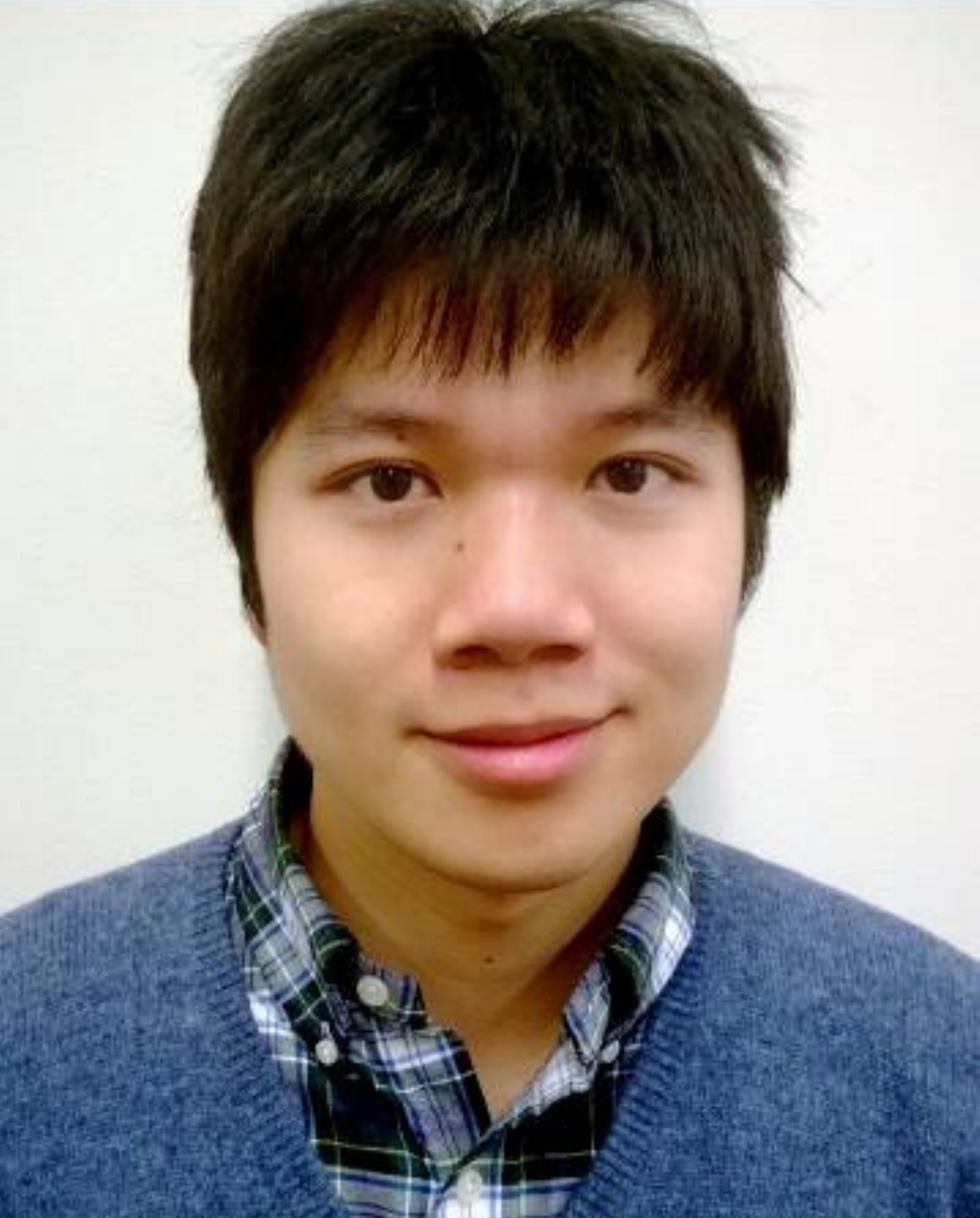}}]{Yingjie Lao} (Senior Member, IEEE) is currently an assistant professor in the Department of Electrical and Computer Engineering at Clemson University. He received the B.S. degree from Zhejiang University, China, in 2009, and the Ph.D. degree from the Department of Electrical and Computer Engineering at University of Minnesota, Twin Cities in 2015. He is the recipient of an NSF CAREER Award, a Best Paper Award at the International Symposium on Low Power Electronics and Design (ISLPED), and an IEEE Circuits and Systems Society Very Large Scale Integration Systems Best Paper Award. 
\end{IEEEbiography}

\begin{IEEEbiography}[{\includegraphics[width=1in,height=1.26in,clip]{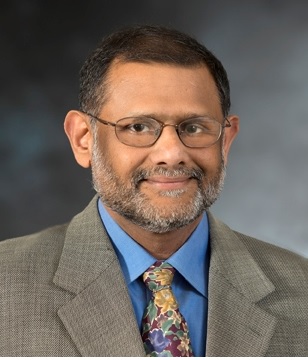}}]{Keshab K. Parhi} (Fellow, IEEE) is the Erwin A. Kelen Chair in Electrical Engineering and a Distinguished McKnight University Professor in the Department of Electrical and Computer Engineering at the university of Minnesota. He completed his Ph.D. in EECS at the University of California, Berkeley in 1988. He has published over 700 papers, is the inventor of 34 patents, and has authored the textbook VLSI Digital Signal Processing Systems (Wiley, 1999). His current research addresses VLSI architectures for machine learning, hardware security, data-driven neuroscience and DNA computing. Dr. Parhi is the recipient of numerous awards including the 2003 IEEE Kiyo Tomiyasu Technical Field Award, and the 2017 Mac Van Valkenburg award and the 2012 Charles A. Desoer Technical Achievement award from the IEEE Circuits and Systems Society. He served as the Editor-in-Chief of the {\em IEEE Trans. Circuits and Systems, Part-I: Regular Papers} during 2004 and 2005. He is a Fellow of the ACM, AIMBE, AAAS, and NAI.
\end{IEEEbiography}
\newpage

% \begin{bibunit}
\setcounter{section}{0}
\setcounter{algorithm}{0}

\title{
Supplementary Information: PaReNTT: Low-Latency Parallel Residue Number System and NTT-Based Long Polynomial Modular Multiplication for Homomorphic Encryption
}

\maketitle

\date{}
\maketitle
\section*{{Supplementary Information: PaReNTT: Low-Latency Parallel Residue Number System and NTT-Based Long Polynomial Modular Multiplication for Homomorphic Encryption}}
This Supplementary Information briefly describes the algorithms for polynomial modular multiplication via the frequency domain using NTT/iNTT. Two approaches are reviewed: negative wrapped convolution (NWC) and low-complexity NWC. A detailed tutorial on this topic is presented in \cite{chiu2023nttbased}.

% \subsection{NTT-based polynomial multiplication using negative wrapped convolution}
An efficient number theoretic transform (NTT)-based polynomial multiplication method with the time complexity of $\mathcal{O}(n\log n)$ is used. This method significantly reduces the time complexity compared to the $\mathcal{O}(n^{2})$ complexity method of the schoolbook polynomial multiplication along with the modular polynomial reduction.

The prior work in~\cite{lyubashevsky2008swifft} presents an efficient algorithm for the NTT-based polynomial multiplication computing $p(x) = a(x) \cdot b(x) \mod (x^n+1,q)$, namely negative wrapped convolution, as  shown in Algorithm \ref{algmpoly}. Note that the weighted operations are needed before NTT and after iNTT during the negative wrapped convolution to avoid the expensive zero padding~\cite{lyubashevsky2008swifft}.  

The core step of this algorithm is the NTT that converts the polynomials $a(x)$ and $b(x)$ to their NTT-domain $\widetilde{A}(x)$ and $\widetilde{B}(x)$ as in Step 2. The NTT for polynomial $a(x)$ is mathematically expressed as 
\begin{equation}
\eqnlbl{ntt}
    \widetilde{A}_k = \sum_{j=0}^{n-1}  a_j\psi^j_{2n} \omega_n^{kj}\mod q,\quad  k \in [0,n-1].
\end{equation}
Polynomial $b(x)$ is similarly transformed to $\widetilde{B}(x)$. Specifically, $\omega$ is the primitive $n$-th root of unity modulo $q$ (i.e., twiddle factor), which satisfies $\omega^{n} \equiv 1 \mod q$. $\psi_{2n}$ is the primitive $2n$-th root of unity modulo $q$, and thus $\omega = \psi_{2n}^2  \mod q$. After using the NTT algorithm, the efficient point-wise multiplication between $\widetilde{A}(x)$ and $\widetilde{B}(x)$ is performed, which is followed by the inverse NTT (iNTT). The iNTT transforms product, $\widetilde{P},$ to the original algebraic domain polynomial $p(x)$, which is defined as
\begin{equation}
\eqnlbl{intt}
    p_k = n^{-1}\psi_{2n}^{-k}\sum_{j=0}^{n-1}  \widetilde{P}_j \omega_n^{-kj}\mod q,\quad  k \in [0,n-1],
\end{equation}
where $n^{-1}$ is the modular multiplicative inverse of $n$ with respect to modulo $q$. 

During the NTT and iNTT, the weighted operation requires the multiplication of the polynomials by the weights $\psi_{2n}^j\mod q$ for NTT or $\psi_{2n}^{-j}\mod q$ for iNTT. Furthermore, an NTT-compatible prime is also utilized, i.e., $q$ must satisfy that $(q-1)$ is divisible by $2n$. 

\begin{algorithm}[htbp]
\caption{\textbf{Negative Wrapped Convolution~\cite{lyubashevsky2008swifft}}}
\label{algmpoly}
\hspace*{\algorithmicindent} \textbf{Input:} $a(x), b(x) \in R_{n,q}$

\hspace*{\algorithmicindent} \textbf{Output:} ${p}(x) = a(x) \cdot b(x) \mod (x^n+1,q)$
 
\begin{algorithmic}[1]
     \STATE Weighted operation:\\
     $\widetilde{a}(x) =\sum_{j=0}^{n-1} a_j\psi_{2n}^jx^j \mod q$ \\
     $\widetilde {b}(x) = \sum_{j = 0}^{n-1} b_j\psi_{2n}^jx^j \mod q$
   \STATE NTT computation: \\
   $\widetilde{A}(x): A_k = \sum_{j=0}^{n-1} \widetilde a_j\omega_{n}^{kj}\mod q$, $ k \in [0,n-1]$ \\ 
   $\widetilde{B}(x): B_k = \sum_{j=0}^{n-1} \widetilde b_j\omega_{n}^{kj}\mod q$, $ k \in [0,n-1]$ 
   \STATE  Point-wise multiplication:\\
   $\widetilde{P}(x) = \widetilde{A}(x) \odot \widetilde{B}(x) = \sum_{k=0}^{n-1} \widetilde {A}_k \widetilde{B}_kx^k$ 
    \STATE iNTT computation:\\
    $\tilde{p}(x) = n^{-1}\sum_{j=0}^{n-1} \tilde P_j\omega_{n}^{-kj}\mod q$, $k \in [0,n-1]$
   \STATE  Weighted operation:\\
   $p(x) =  \sum_{j=0}^{n-1}  \widetilde{p_j}  \psi_{2n}^{-j}x^j$
\end{algorithmic}
\end{algorithm}

Since the weighted operations in NTT/iNTT require a large number of expensive modular multiplications, the recent works in~\cite{zhang2020highly,longa2016speeding} present a new method to merge the weighted operations into the butterfly operations. In particular, the new NTT in \eqnref{ntt} is re-represented as $ \widetilde{A}_k$ and $ \widetilde{A}_{k+n/2}$ by using the decimation-in-time (DIT) method:
\begin{align}
    \widetilde{A}_k &= a^{(0)}_k + \psi_{2n} \omega^k_n a_k^{(1)} \mod q \eqnlbl{ntt_merged_upper}, \\
     \widetilde{A}_{k+n/2} &= a^{(0)}_k - \psi_{2n} \omega^k_n a_k^{(1)} \mod q \eqnlbl{ntt_merged_lower},
\end{align}
where $k\in [0,\frac{n}{2}-1]$ and 
\begin{align}
    a^{(0)}_k &=  \sum_{j=0}^{n/2-1}  a_{2j}\psi^j_{n} \omega_{n/2}^{kj}\mod q,\\
    a^{(1)}_k &= \sum_{j=0}^{n/2-1}  a_{2j+1}\psi^j_{n} \omega_{n/2}^{kj}\mod q.
\end{align}
Since $\omega = \psi_{2n}^2  \mod q$, integers $\psi^j_{2n}$ and $\omega_{n/2}^{kj}$ can be merged as an integer $\psi_{2n} \omega^k_n = \psi^{(2k+1)}_{2n}$. Thus, only one modular multiplication is required in the butterfly operation.

The improved iNTT algorithm merges not only the weighted operation but also the multiplication with constant $n^{-1}$ into the butterfly operations, as presented in~\cite{zhang2020highly}. Based on \eqnref{intt} and the decimation-in-frequency (DIF) method, the new iNTT algorithm is expressed as 
\begin{align}
    p_{2k} &= (\frac{n}{2})^{-1}  \psi^{-k}_{n} \sum_{j=0}^{n/2-1}  \widetilde{P}^{(0)}_j \omega^{-kj}_{n/2} \mod q, \eqnlbl{intt_merged_even} \\
    p_{2k+1} &= (\frac{n}{2})^{-1} \psi^{-k}_{n}\sum_{j=0}^{n/2-1}   \widetilde{P}^{(1)}_j \omega^{-kj}_{n/2} \mod q \eqnlbl{intt_merged_odd},
\end{align}
where $k\in [0,\frac{n}{2}-1]$, and 
\begin{align}
    \widetilde{P}^{(0)}_j &= \frac{\widetilde{P}_j+\widetilde{P}_{j+n/2}}{2} \mod q \\
    \widetilde{P}^{(1)}_j &= \frac{\widetilde{P}_j-\widetilde{P}_{j+n/2}}{2} \omega_n^{-j}\psi_{2n}^{-1} \mod q.
\end{align}
Similarly, the integers $\omega_n^{-j}$ and $\psi_{2n}^{-1}$ are combined as an integer $\psi^{-1}_{2n} \omega^{-j}_n = \psi^{-(2j+1)}_{2n}$. Different from the NTT butterfly architecture, the modular addition and modular subtraction intermediate results in the iNTT butterfly need to be divided by two. In fact, the modular multiplication by $2^{-1}$ can be implemented without a modular multiplier~\cite{zhang2020highly}, since
\begin{align}
    \frac{x}{2} = 
    \begin{cases}
        \frac{x}{2},& \text{if } x \text{ is even} \\
        \lfloor \frac{x}{2} \rfloor + \frac{q+1}{2} \mod{q},& \text{if } x \text{ is odd}.
    \end{cases}
\end{align}

Specifically, $x \times 2^{-1}$ can be implemented as $(x \gg 1)$ when $x$ is even. If $x$ is odd, $x \times 2^{-1}$ can be represented as:
\begin{align}
    \frac{x}{2} &\equiv (2 \lfloor \frac{x}{2}\rfloor + 1)\frac{q+1}{2} \nonumber \\
                &\equiv  \lfloor \frac{x}{2} \rfloor (q+1) + \frac{q+1}{2} \nonumber \\
                &\equiv  \lfloor \frac{x}{2} \rfloor + \frac{q+1}{2} \mod{q} \label{eq:two_inv}
\end{align}
where $\lfloor \frac{x}{2} \rfloor$  can also be implemented as $(x \gg 1)$, and $(q+1)/2$ is a pre-computed constant. Hence, no modular multiplications are required while requires one modular adder and a multiplexer (MUX). 
% \putbib
% \end{bibunit}
% \bibliography{refs}

% \input{response}
\end{document}